\documentclass [sigconf]{acmart}
\AtBeginDocument{%
  }

\copyrightyear{2026}
\acmYear{2026}
\setcopyright{cc}
\setcctype{by-nc-nd}
\acmConference[CHI '26]{Proceedings of the 2026 CHI Conference on Human Factors in Computing Systems}{April 13--17, 2026}{Barcelona, Spain}
\acmBooktitle{Proceedings of the 2026 CHI Conference on Human Factors in Computing Systems (CHI '26), April 13--17, 2026, Barcelona, Spain}
\acmDOI{10.1145/3772318.3790830}
\acmISBN{979-8-4007-2278-3/2026/04}

\usepackage{url} 

\usepackage{booktabs}
\usepackage[normalem]{ulem}
\useunder{\uline}{\ul}{}
\usepackage{longtable}

\usepackage{tabularx}
\usepackage{booktabs}
\usepackage{ragged2e} 

\newcommand{\Revision}[1]{{\color{black}#1}}

\begin{document}

\title{From Pets to Robots: MojiKit as a Data-Informed Toolkit for Affective HRI Design}

\author{Liwen He}
\orcid{0000-0003-0715-9252}
\affiliation{%
    \institution{The Hong Kong University of Science and Technology (Guangzhou)}
    \city{Guangzhou}
    \state{Guangdong}
    \country{China}
    \institution{Tsinghua University}
    \city{Beijing}
    \state{}
    \country{China}}
\email{helw24@mails.tsinghua.edu.cn}

\author{Pingting Chen}
\orcid{0009-0000-9442-1559}
\affiliation{%
    \institution{The Hong Kong University of Science and Technology (Guangzhou)}
    \city{Guangzhou}
    \state{Guangdong}
    \country{China}
    \institution{Beijing Institute of Technology}
    \city{Beijing}
    \state{}
    \country{China}}
\email{3120231887@bit.edu.cn}

\author{Ziheng Tang}
\orcid{0009-0001-9612-0300}
\affiliation{%
    \institution{The Hong Kong University of Science and Technology (Guangzhou)}
    \city{Guangzhou}
    \state{Guangdong}
    \country{China}
    \institution{Tsinghua University}
    \city{Beijing}
    \state{}
    \country{China}}
\email{tgzihg@gmail.com}

\author{Yixiao Liu}
\orcid{0009-0001-9612-0300}
\affiliation{%
    \department{}
    \institution{Sichuan University}
    \city{Chengdu}
    \state{Sichuan}
    \country{China}}
\email{liuyixiao1@stu.scu.edu.cn}

\author{Jihong Jeung}
\orcid{0009-0005-8925-0712}
\affiliation{%
    \department{}
    \institution{Tsinghua University}
    \city{Beijing}
    \state{}
    \country{China}}
\email{jihong95@mail.tsinghua.edu.cn}

\author{Teng Han}
\orcid{0000-0001-8857-8787}
\affiliation{%
    \department{Institute of Software,}
    \institution{Chinese Academy of Sciences}
    \city{Beijing}
    \state{}
    \country{China}}
\email{hanteng1021@gmail.com}

\author{Xin Tong}
\authornote{Corresponding author.}
\orcid{0000-0002-8037-6301}
\affiliation{%
    \institution{The Hong Kong University of Science and Technology (Guangzhou)}
    \city{Guangzhou}
    \state{Guangdong}
    \country{China}
    \department{}
    \institution{The Hong Kong University of Science and Technology}
    \city{Hongkong}
    \state{}
    \country{China}}
\email{xint@hkust-gz.edu.cn}

\renewcommand{\shortauthors}{He et al.}

\begin{abstract}
  Designing affective behaviors for animal-inspired social robots often relies on intuition and personal experience, leading to fragmented outcomes. To provide more systematic guidance, we first coded and analyzed human–pet interaction videos, validated insights through literature and interviews, and created structured reference cards that map the design space of pet-inspired affective interactions. Building on this, we developed MojiKit, a toolkit combining reference cards, a zoomorphic robot prototype (MomoBot), and a behavior control studio. We evaluated MojiKit in co-creation workshops with 18 participants, finding that MojiKit helped them design 35 affective interaction patterns beyond their own pet experiences, while the code-free studio lowered the technical barrier and enhanced creative agency. Our contributions include the data-informed structured resource for pet-inspired affective HRI design, an integrated toolkit that bridges reference materials with hands-on prototyping, and empirical evidence showing how MojiKit empowers users to systematically create richer, more diverse affective robot behaviors.


\end{abstract}



\begin{teaserfigure}
  \includegraphics[width=\textwidth]{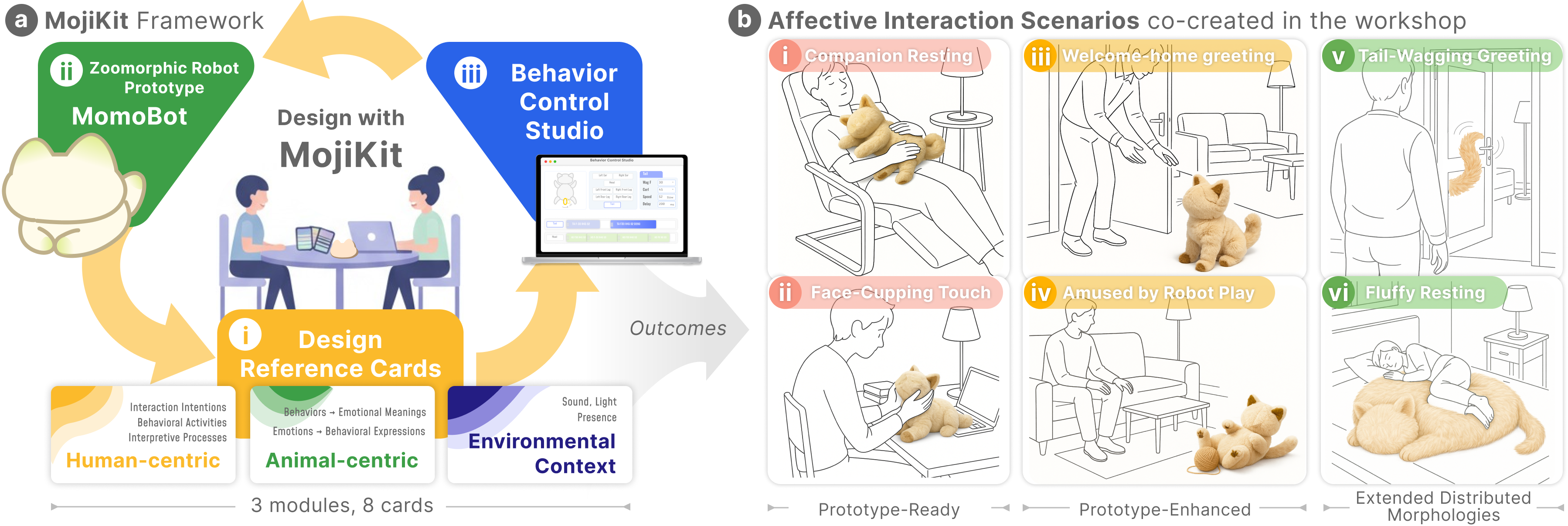}
  \caption{Overview of the MojiKit.
(a) MojiKit integrates a zoomorphic robot prototype (MomoBot), design reference cards (human-, animal-centric, and environmental context), and a behavior control studio, to support emotion-oriented interaction design. (b) Affective interaction scenarios co-created in workshops, organized into three levels of realizability: Directly realizable with the current prototype (Companion Resting, Face-Cupping Touch), Enabled by moderate enhancements (Welcome-home Greeting, Amused by Robot Play), 
and Extended distributed morphologies (Tail-Wagging Greeting, Fluffy Resting).
}
  \Description{Figure shows the MojiKit framework and example design outcomes. Panel (a) illustrates three components: the zoomorphic robot prototype “MomoBot,” the Behavior Control Studio, and the Design Reference Cards, organized into human-centric, animal-centric, and environmental modules. Panel (b) depicts six affective interaction scenarios co-created in workshops: (i) companion resting, (ii) face-cupping touch, (iii) welcome-home greeting, (iv) amused by robot play, (v) tail-wagging greeting, and (vi) fluffy resting. Scenarios are positioned along a continuum from directly realizable with the prototype, to requiring moderate enhancements, to extended distributed morphologies. }
  \label{fig:teaser}
\end{teaserfigure}


\maketitle

\section{Introduction}

Pets contribute to human well-being by offering psychological comfort, emotional regulation, and social support \cite{allen2003pets,fine2010handbook,allen2001pet}. Prior work shows that animal-inspired robots can provide similar benefits in a range of contexts. 
\Revision{Robotic pets such as PARO \cite{wada2007living}, Joy for All \cite{schnitzer2024prototyping}, MiRo \cite{htet2025hey}, and AIBO \cite{fujita2001aibo} have been found to reduce loneliness and stress among older adults and residents in long-term care \cite{wada2004effects,brodie1999exploration}, and support socio-emotional engagement for children, including those with developmental needs \cite{hedayati2019hugbot,melson2009robotic,paluch2022s}. Beyond therapeutic contexts, the capacity of such robots to provide emotional resonance extends to everyday environments \cite{yoshida2022production}, offering companionship to general users whose living or health conditions make real pet ownership difficult \cite{koivusilta2006have,baxter1984deleterious}.


Zoomorphic robots, which model their social behavior on animals, rely heavily on motion and non-verbal cues for human-centric communication \cite{fong2003survey,breazeal2003emotion}. Yet, designing non-verbal expressive behaviors remains a challenge, as strict biological mimicry can produce subtle movements that are realistic but may lack the clarity required for user interpretation \cite{mancini2011animal}. A robot’s interaction is experienced as enjoyable and competence-enhancing if and only if it adheres to human social expectations \cite{reeves1996media,fong2003survey}. Therefore, effective companion robot behaviors should balance biological plausibility with social readability to support affective communication \cite{breazeal2003emotion}. 
However, current approaches typically involve either rigid hard-coding or complex technical workflows \cite{quigley2009ros,macenski2022robot,hussain2025low}. This highlights a lack of systematic and accessible tools, especially for non-expert users, to explore or customize expressive motions for animal-inspired robots.



}

Animal-inspired robots can build on users’ natural tendency to attribute emotions and intentions to animal-like behaviors. However, current design approaches provide limited support for systematically modeling and implementing these interpretations. Broadly, existing work falls into two categories. The first approach pursues high-fidelity replication of animal repertoires using ethological models and data-driven techniques \cite{miklosi2017ethorobotics}. While this can reproduce realistic movements, it does not identify which behaviors are most relevant for emotional communication, and often proves less effective in companionship contexts where subtle cues are more critical than realism \cite{mori2012uncanny}. A second, and often more suitable, approach for social HRI is intuition-driven design, in which users focus on salient gestures (e.g., tail wagging, ear tilt, gaze, posture shifts) based on how people naturally interpret affective information in animals \cite{macdonald2024evaluating, yohanan2012role, gacsi2016humans}. However, this route relies heavily on intuition and personal experience \cite{jung2017affective}, typically resulting in isolated behavior fragments and lacking systematic, data-informed references to link behaviors with emotional meaning or to support composing them into coherent interactions \cite{ghafurian2022zoomorphic, sanoubari2022from}. Moreover, even when users wish to explore the expressive potential of different behaviors, they lack flexible, hands-on tools for organizing, combining, and prototyping behaviors in practice. As a result, opportunities to experiment with nuanced emotional interactions and to iteratively refine design ideas remain limited. Current challenges therefore span both knowledge resources and practical tools.

To address these challenges, we investigate how to better support the creative, intuition-driven design of animal-inspired social robots. Our first research question is: \textbf{RQ1. Which affective behavior patterns observed in human--pet interaction can inform the emotional interaction design of animal-inspired social robots?} To explore this question, we developed \textbf{MojiKit}, a toolkit that preserves the value of users’ personal intuition while improving scalability and reusability. MojiKit consists of two components: (1) a structured design reference that makes the design space of affective behaviors explicit, and (2) a prototyping platform that enables rapid implementation, testing, and refinement of behaviors in real time. Using MojiKit as a design probe, we further ask: \textbf{RQ2. How do users use a data-informed toolkit to conceptualize and prototype emotionally expressive interaction patterns, and what insights emerge for the design of affective social robots?}

To build \textit{MojiKit}, we analyzed 65 video clips of human-pet interactions and validated them against ethological literature to derive a set of structured \textbf{Design Reference Cards}. We then developed a tangible prototype (\textbf{MomoBot}) and a graphical interface (\textbf{Behavior Control Studio}) (see Figure~\ref{fig:teaser}), together forming an exploratory design probe that enables users to retrieve, combine, and sequence animal-like behaviors into expressive interaction patterns. We evaluated MojiKit through pair-wise co-creation workshops with 18 participants. Across the sessions, participants generated 35 affective interaction patterns, demonstrated how simple primitives can be combined into complex sequences, and revealed trajectories between behavioral realism and human-centered emotional needs. The findings also highlight opportunities to extend emotional expression through distributed morphological designs embedded in everyday environments.  

This work advances HRI in three ways. It formalizes an actionable schema for affective behavior design to enable systematic comparison, reuse, and teaching across projects, by linking intents, triggers, pet (robot) motion primitives, and human interpretations. It establishes a practical pathway for intuition-driven design by coupling structured guidance with embodied, real-time iteration, lowering entry barriers and accelerating exploration for non-experts. Finally, it contributes concrete strategies and trajectories for zoomorphic companion robots, providing targeted design guidance and future priorities for affective social robot design. \Revision{To facilitate further research and community adoption, we contribute the MojiKit platform (including hardware schematics, software, and design resources) as an open-source toolkit \footnote{\url{https://github.com/ARKxLab/MojiKit}}.}

\section{Related Work}
\subsection{\Revision{Human Interpretation in Human-Pet and Human-Robot Affective Bonding}}

The emotional connection between humans and pets has been extensively studied through the lens of \textit{attachment theory} \cite{topal1998attachment}. 
While biological synchrony (e.g., oxytocin levels) plays a role \cite{nagasawa2015oxytocin}, this bond is fundamentally shaped by human cognition. 
Owners actively engage in \textbf{anthropomorphism}, attributing human-like emotions and intentions to animal behaviors based on perceived responsiveness \cite{epley2007seeing, serpell1996company}. 
Since an animal’s internal state is inaccessible, the subjective ``meaning-making'' process transforms raw animal movements into shared emotional experiences \cite{horowitz2009disambiguating}. \Revision{This reliance on interpretation extends to human-robot interaction, where animal-like features invite users to project similar social schemas onto mechanical agents \cite{fong2003survey}.
According to the \textit{Computers Are Social Actors (CASA)} paradigm \cite{nass1994computers}, humans tend to apply social rules and models to artifacts that exhibit animate cues \cite{reeves1996media}. 
When interacting with a zoomorphic robot, users naturally attempt to interpret its motions through the same lens they use for living pets, seeking intent and emotion in mechanical actions \cite{lee2006can,breazeal2004designing}. 
Therefore, the formation of an emotional bond requires a synergy: the robot's behaviors should be grounded in biological agency \cite{bates1994role}, yet ensure sufficient \textbf{readability} to solicit and sustain the user's subjective interpretation \cite{takayama2011expressing, breazeal2003emotion}. 
This suggests a design approach that prioritizes perceptual validity over mechanical replication. Rather than relying solely on ethological models, effective design requires identifying the specific behavioral cues that trigger emotional attribution in human observers \cite{takayama2011expressing}. This motivates MojiKit’s methodology: systematically capturing human interpretations of animal behaviors to inform the design of expressive robot primitives.}

\Revision{
\subsection{Animal–Computer Interaction and More-than-Human Perspectives}

As zoomorphic robots increasingly operate in everyday settings, Animal–Computer Interaction (ACI) and more-than-human HCI provide a critical multispecies perspective, framing these technologies as active participants in shared ecologies rather than isolated tools \cite{mancini2013animal,westerlaken2024design}.
Beyond technologies explicitly built for animals, such as enrichment tools \cite{schneiders2024designing,mancini2015re} or mediated communication platforms \cite{kleinberger2023birds,xue2024dogchat}, recent ACI scholarship emphasizes that interactive systems are deeply embedded in multispecies ecologies \cite{westerlaken2025designing,kleinberger2025animals}. Researchers have explored this by designing environments shared by robots, cats, and humans \cite{westerlaken2024matters} and analyzing the ethical tensions of technological interventions \cite{westerlaken2020imagining}. These inquiries suggest that, to be truly animal-centric, design decisions must be informed by ethological data rather than mere projection \cite{schneiders2023tas}. ACI scholarship also raises concerns about technological substitution and its potential consequences for human–animal relationships and animal welfare~\cite{sparrow2002march,sharkey2020crying}.

Therefore, we developed the animal-centric component of our \textbf{Design Resource Cards} based on ethological literature  (Section \ref{sec:Literature Integration}), ensuring that behavior design remains informed by the natural patterns of real animal behavior.

}

\subsection{Designing Emotionally Expressive Behaviors in Zoomorphic Robots}

Zoomorphic robots are increasingly used to study emotional companionship and social interaction in HRI \cite{schnitzer2024prototyping}. A key challenge is creating emotionally expressive behaviors that users can intuitively interpret. Prior work generally follows two approaches: data-driven replication and intuition-driven design. \textbf{Data-Driven Replication.} This approach seeks high-fidelity reproduction of animal motions using ecological and ethological models \cite{miklosi2017ethorobotics}. Techniques such as motion capture, biomechanics modeling, and deep learning capture lifelike gestures, as seen in Boston Dynamics’ robotic dogs\footnote{\url{https://bostondynamics.com/products/spot/}} or projects that learn animal sequences from video and sensor data. While these systems achieve impressive realism, they are costly and risk uncanny valley effects in social contexts \cite{mori2012uncanny}. Moreover, they emphasize biomechanical fidelity over emotional readability, limiting suitability for long-term affective interaction \cite{sauer2021zoomorphic}. \textbf{Intuition-Driven Design.} Alternatively, designers selectively replicate salient animal actions based on emotional interpretation, focusing on gestures that signal specific affective states (e.g., tail wagging, gaze, ear movements) \cite{wang2024enhancing, sauer2021zoomorphic, gacsi2016humans}. The Haptic Creature \cite{yohanan2012role}, for instance, mimics breathing rhythms and purring to evoke comfort. Such methods often yield approachable and emotionally expressive robots, aligning with human tendencies for affective projection \cite{sefidgar2015design}. However, intuition-driven design is subjective and difficult to replicate or scale \cite{miklosi2017ethorobotics}, and most works remain limited to isolated behavioral fragments (e.g., wagging tails, curling postures) without a systematic framework linking behavior to affective meaning \cite{macdonald2024evaluating, ghafurian2022zoomorphic}.
 
Both approaches have advanced the field but lack systematic means to design emotionally coherent, complex patterns. Few resources help designers organize and combine behaviors into interactive sequences that align with human interpretations of emotion, personality, and social intention \cite{sanoubari2022from}. Our work addresses this gap by introducing a structured, reusable behavior dataset and toolkit to support intuitive yet scalable emotional behavior design for zoomorphic robots.

\begin{figure*}
    \centering
    \includegraphics[width=0.95\linewidth]{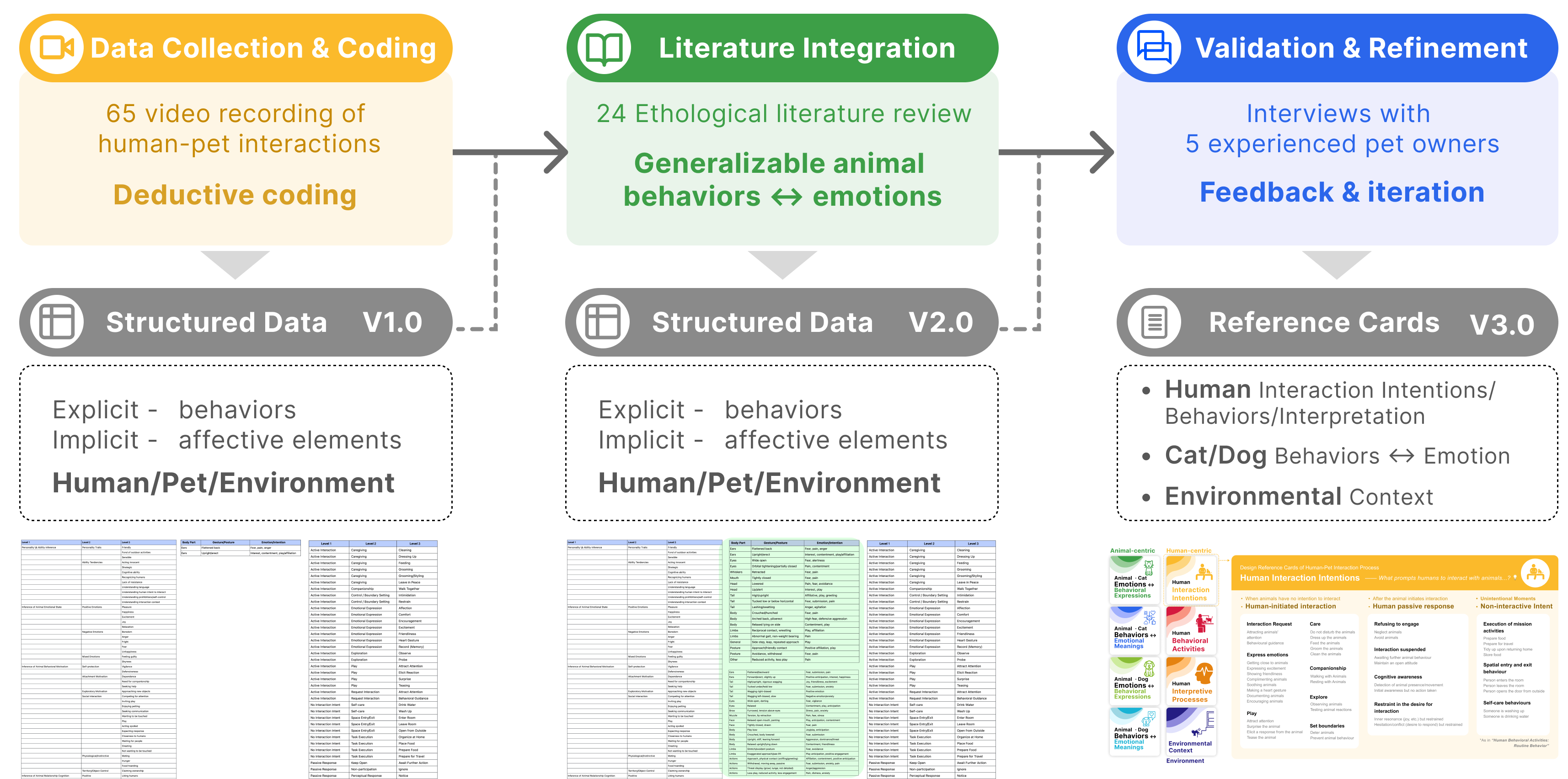}
    \caption{Design process of the reference cards: from video-based deductive coding (V1.0), to literature integration on animal behaviors and emotions (V2.0), to refinement via interviews with pet owners, resulting in the finalized \textbf{Design Reference Cards V3.0}.}
    \Description{Figure illustrates the design process of the reference cards in three stages. Left: data collection and deductive coding of 65 human–pet interaction videos, producing Structured Data V1.0 with explicit behaviors and implicit affective elements across human, pet, and environment. Center: integration of 24 ethological studies, adding generalizable animal behaviors and emotion mappings, resulting in Structured Data V2.0. Right: validation and refinement through interviews with five pet owners, producing Reference Cards V3.0 organized into three modules, eight cards: human intentions/behaviors/interpretations, cat–dog behavior–emotion mappings, and environmental context. }
    \label{fig:design process of reference cards}
\end{figure*}

\subsection{The Design Features of Animal-inspired Companion Robots}
Animal-inspired robots use multimodal strategies to convey emotion, combining appearance, facial expressions, sound, and behavior \cite{yoshida2022production,flagg2013affective,li2025sofibuddy}. Likeability is strongly shaped by appearance \cite{loffler2020uncanny}, with designs ranging from hard resin shells (e.g., AIBO) \cite{fujita2001aibo} to plush surfaces that simulate petting \cite{flagg2013affective}. Facial expressions are created through mechanical panels or screen displays \cite{zhang2014ebear,thomas2025neffy}, but overly realistic or awkward designs risk uncanny effects \cite{loffler2020uncanny}. Robots such as Lovot combine facial changes with vocalizations, enhancing trust through multimodal consistency \cite{tsiourti2019multimodal}. However, ambiguous onomatopoeia \cite{komatsu2015choreographing} can reduce clarity, and overly advanced LLM-based speech can lower acceptance \cite{kim2024understanding}. Interactive behaviors are central to forming emotional bonds in both human–pet and human–robot relationships \cite{topal1998attachment,farago2014social}. Simple movements, such as approaching or retreating, can signal closeness or distance and shape users’ feelings of attachment \cite{yonezawa2024snuggling}. Beyond isolated actions, animal-inspired robots often rely on coordinated whole-body movements, which enhance the richness and credibility of emotional expression \cite{tsiourti2019multimodal,li2024emopus}. For instance, when Lovot seeks a hug, it synchronizes head tilts, forward leaning, and rhythmic arm waving to evoke  perceived naturalness and warmth \cite{yoshida2022production}. Sustained physiological cues, such as breathing rhythms, can further communicate arousal levels and internal states, strengthening users’ perception of the robot’s emotional presence \cite{yohanan2011design}.

Despite these advances, most designs rely on fixed presets, limiting openness and personalization. Users cannot flexibly combine body parts or tailor behaviors to emotional goals. To address this, our work introduces an open-source prototyping platform that integrates hardware and software, supports code-free customization, and embeds human-pet interaction pattern knowledge to help users map animal behaviors to human emotional interpretations, enabling scalable and expressive interaction design.

\section{Construction of the Design Resource}

Our goal was to develop a systematic reference resource that represents the design space of emotional interaction, thereby supporting the intuitive design of animal-inspired robots. Rather than relying solely on designers’ personal experiences, this design space is intended to help them recall and consider possible patterns of human–pet interaction.

We collected and analyzed video recordings of human–pet interactions, using observational coding to capture the dynamics of emotional exchange. Guided by the input–output interaction paradigm, we structured the process into three categories (human, animal, and environment), each divided into explicit behaviors and implicit affective elements such as interpretations and motivations. To ensure generalizability, animal behaviors and emotional meanings were cross-validated with ethological literature. The structured resource was then iteratively refined through interviews with five experienced pet owners, resulting in a set of reference cards for designing affective interactions (see Figure~\ref{fig:design process of reference cards}).

\subsection{Video Analysis of Human–Pet Interaction}
To obtain naturalistic records of everyday human–pet interaction, we collected publicly available videos from major social media platforms, focusing on cats (Felis catus) and dogs (Canis familiaris). Using Chinese keywords and their colloquial variants, we retrieved 246 clips with over 1,000 “likes.” For each candidate, we documented the URL, platform, interaction metrics, duration, and presence of narration or on-screen text. 

We screened candidate videos based on the following criteria: inclusion required spontaneous, emotionally oriented daily interactions with visible pet behaviors and human reactions, supported by verbal or textual cues; exclusion applied to staged or edited content, task-oriented training scenes, anthropomorphic exaggerations, and low-quality footage. After screening, 65 videos were retained for analysis\Revision{, totaling approximately 79 minutes of footage ($M=72.48s$, $SD=101.88s$). This variation in clip length allowed us to capture both instantaneous affective signals in short clips and sustained interaction dynamics in longer sequences.}

\subsubsection{Development of the Coding Scheme}

We conducted a structured deductive analysis \cite{azungah2018qualitative} of the 65 collected videos. An initial draft \textit{codebook (R1)} was developed based on the input--output interaction paradigm. Each video was segmented into interaction rounds with three components (trigger, response, and closure) potentially spanning multiple exchanges. Verbal narration or on-screen text was used as supporting evidence of human interpretation, while viewer comments were considered only when directly aligned with observable behaviors. The draft codebook comprised three dimensions: (1) \textbf{Context:} setting, environment, temporal cues, and external stimuli. (2) \textbf{Human:} interaction intentions, psychological interpretations (e.g., attributions, feelings, planning), and observable behaviors. (3) \textbf{Pet:} animal actions, including whole-body movements and local gestures (e.g., tail/ear positions, limb postures). Coding was performed at the level of minimal identifiable action units (e.g., ``tail tip flick,'' ``ear flattening,'' ``crouching with hindquarter movement''). Each round began with a clear initiating event (e.g., reaching out, pet approaching, or an environmental stimulus), was followed by reciprocal responses, and ended once the interaction stabilized or shifted to a new event.

\subsubsection{Coding Process}

Four researchers contributed to the development of the coding scheme.  
In the first stage, three researchers independently coded five videos using codebook \textit{R1}, then 
discussed discrepancies in segmentation, granularity, and attribution rules. This resulted in an updated 
\textit{codebook (R2)} with refined operational definitions and boundary rules.
\Revision{
To further validate and stabilize the scheme, the same three researchers jointly applied \textit{R2} to an 
additional set of 10 videos. The outcomes informed a final round of consolidation, yielding 
\textit{codebook (R3)}. 
To evaluate the reliability of \textit{codebook (R3)}, two independent researchers (one original coder and one new coder) coded a sampled subset of 15 videos (23.08\% of the dataset).  
Inter-rater reliability was calculated using Cohen’s Kappa, resulting in $\kappa = 0.929$, indicating 
substantial agreement~\cite{landis1977measurement}. Differences were resolved through discussion, and the 
agreed definitions were adopted for the remaining dataset.}
The final coding produced 65 multi-turn human--pet interaction sequences. Coded elements were organized into a structured dataset that separated explicit behaviors from implicit affective elements across human, animal, and environmental dimensions. This table (\textbf{Design Reference v1.0}) served as the foundation for constructing the design space.

\begin{figure*}
    \centering
    \includegraphics[width=0.95\linewidth]{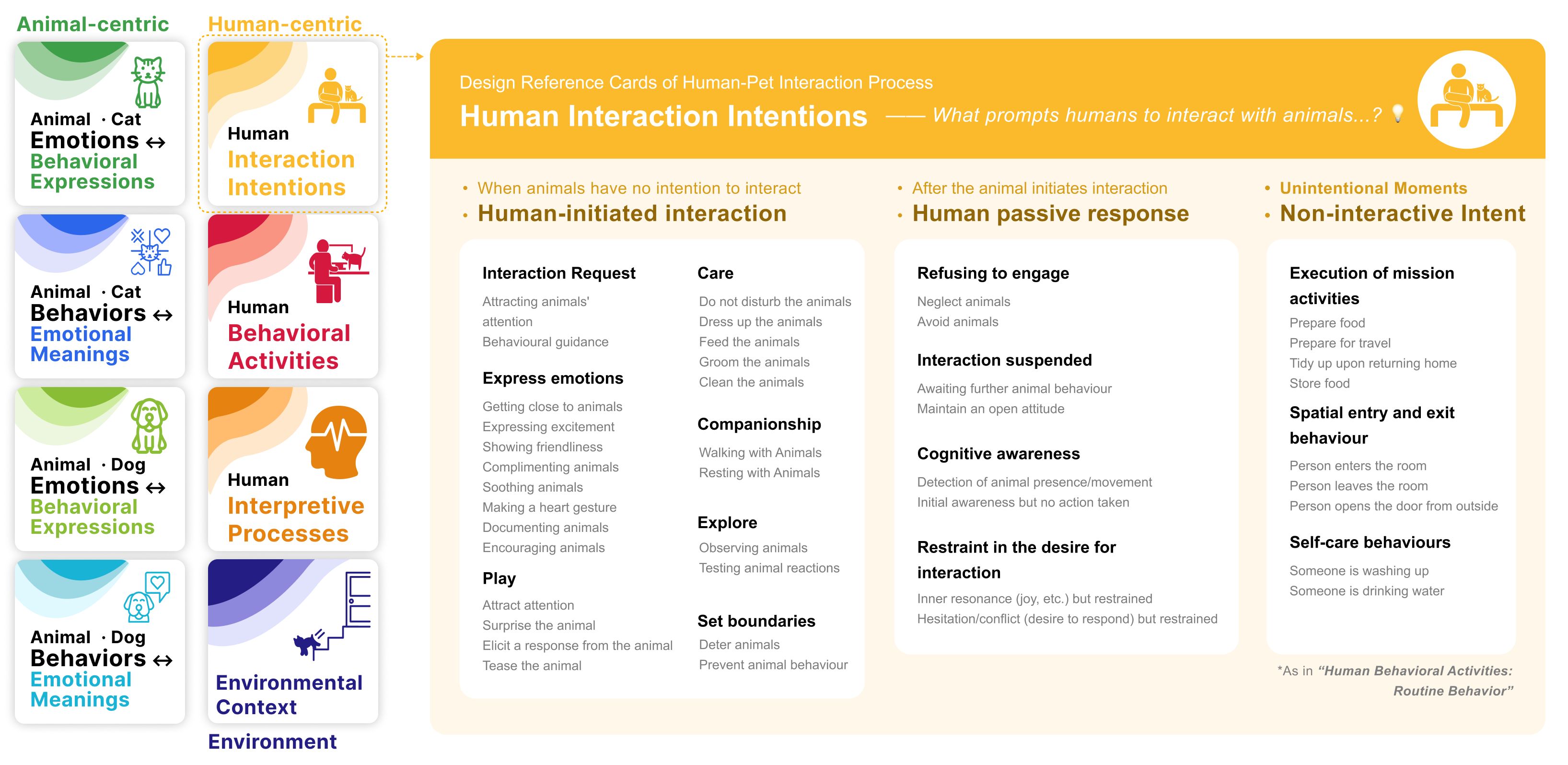}
    \caption{Design Reference Cards comprising three modules (eight cards in total) of human–pet interaction. Example: the “Human Interaction Intentions” card, detailing motives such as human-initiated actions, passive responses, and non-interactive intent.}
    \Description{A two-panel reference card for human-pet interaction design. Left: Color-coded sections categorize animal-centric (e.g., cat or dog emotional expressions), human-centric (e.g., interaction intentions), and environmental cards. Right: Detailed breakdown of human intentions, including proactive interaction, passive response to animal-initiated actions, non-interactive scenarios (e.g., refusal/neglect), and routine behaviors like preparing food or spatial transitions. }
    \label{fig:reference cards}
\end{figure*}

\subsection{Literature Integration}\label{sec:Literature Integration}

To complement the video-based analysis, we reviewed ethological literature on cats (\textit{Felis catus}, $n=15$) and dogs (\textit{Canis familiaris}, $n=9$), focusing on taxonomies and ethograms that link observable signals (e.g., postures, facial configurations, ear/tail positions, whole-body movements) to emotional or motivational states~\cite{evangelista2019facial,siniscalchi2018communication}. We included studies with operationalized definitions and validity evidence, and excluded those limited to acoustic/physiological measures, breed standards, or descriptive lists. For cats, representative sources included CatFACS~\cite{kovavcivc2024cat}, grimace scales, ethograms of affiliative play, and defensive/fearful behaviors~\cite{gajdovs2023ethological}. For dogs, we drew on DogFACS~\cite{waller2013dogfacs} and studies mapping postural/facial units to valence–arousal, as well as work on frustration, anticipation, affiliation, and communicative functions of tail/ear movements~\cite{siniscalchi2018communication,boneh2022explainable,kiley1976tail}. These materials supplemented the animal-centric references (\textit{Behaviors \& Emotional Meanings}). This integration yielded \textbf{Design Reference v2.0}.

\subsection{Evaluation and Iteration}

To assess the clarity, usability, and coverage of the initial design space, we conducted semi-structured interviews with five participants (all female; $M_{\text{age}}=25.8$, $SD=0.83$) who had over three years of pet-keeping experience. Participants were introduced to the purpose of the design space as a source of inspiration for interaction design and then guided through the modules to provide feedback on overall impression, clarity of classification, content coverage, and usability (see interview framework in Appendix \ref{sec:cards design interview}).

The interviews revealed several issues: (1) the table-based format obscured hierarchical relations; (2) the animal modules were difficult to navigate from an emotion-first perspective; and (3) some terms were overly technical due to reliance on ethology sources. Based on this feedback, we revised the design space into card-based references with clearer layout and color coding, offered two complementary views in animal-centric module (\textit{Behaviors $\rightarrow$ Emotional Meanings}, \textit{Emotions $\rightarrow$ Behavioral Expressions}), simplified terminology, and added new entries (e.g., feline hunting behaviors) cross-validated with literature. These refinements resulted in the \textbf{Design Reference Cards v3.0}, which were used in subsequent stages of the study.

\subsection{Design Reference Cards}\label{sec:reference cards}

The final version of the Design Reference Cards integrates the structured data and subsequent iterations into three interconnected modules comprising a total of eight cards (see Figure \ref{fig:reference cards}). Each card addresses a distinct dimension of human–pet interaction: human intentions, behaviors and interpretations; environmental factors; animal behaviors and emotional meanings (separately for cats and dogs). The animal-centric cards are organized to allow both behavior-first and emotion-first entry points, reflecting the dual perspectives embedded in the design. This modular organization provides a systematic yet flexible structure that supports varied design processes. The specific modules are as follows (details in Appendix \ref{sec:8 cards}):

\begin{itemize}
    \item \textbf{Human-Centric Module:}  
    \begin{itemize}
        \item \textbf{Human Interaction Intentions:} outlines motivations from actively seeking engagement to passively observing.  
        \emph{Example: A child waves a teaser toy to attract a kitten’s attention (human-initiated interaction).}

        \item \textbf{Human Behavioural Activities:} provides a taxonomy of human actions, including physical contact, gestures, and communication through voice or objects.  
        \emph{Example: A dog owner bends down to pick up the puppy while softly calling its name, combining direct physical contact with vocal interaction.}

        \item \textbf{Human Interpretations:}  
        Captures the cognitive and emotional processes by which humans interpret animal signals, including attributions of traits, states, and motivations.  
        \emph{Example: When a cat lies on the owner’s lap, the owner interprets it as seeking comfort.}
        \end{itemize}

    \item \textbf{Environmental Module:}  
    Represented by the card \textbf{Environmental Factors}, which highlights external stimuli, such as sounds, light, or the presence of others, that can trigger or disrupt interactions.  
    \emph{Example: A doorbell rings and the dog rushes to the entrance barking.}

    \item \textbf{Animal-Centric Module:}  
    Divided into two parallel sets of cards for cats and dogs, each offering both behavior-first and emotion-first perspectives.  
    \begin{itemize}
        \item \textbf{Dog Emotions \& Corresponding Behaviors} — lists behaviors associated with affective states such as joy, fear, or aggression.  
        \emph{Example: A puppy wags its tail and barks excitedly, expressing joy and playfulness.}

        \item \textbf{Dog Behaviors \& Emotional Meanings} — organizes actions by body part (e.g., ears, tail, body posture) and links them to inferred emotions.  
        \emph{Example: A puppy flattens its ears and crouches low, indicating fear or submission.}

        \item \textbf{Cat Emotions \& Corresponding Behaviors} — presents behaviors linked to affective states like playfulness, anger, or contentment.  
        \emph{Example: A kitten rolls on its back and exposes its belly, often interpreted as playful joy.}

        \item \textbf{Cat Behaviors \& Emotional Meanings} — organizes gestures (ears, whiskers, body, tail) and their affective interpretations.  
        \emph{Example: A kitten retracts its whiskers and stiffens its body, signaling pain or high stress.}
    \end{itemize}
\end{itemize}

By offering interconnected yet independent modules, the card set forms a multi-facet structure that goes beyond a linear cause–effect model. It enables users to explore feedback loops where animal expressions trigger human interpretations, supporting both systematic inquiry and open-ended ideation.

\section{MojiKit}

\begin{figure*}
    \centering
    \includegraphics[width=\linewidth]{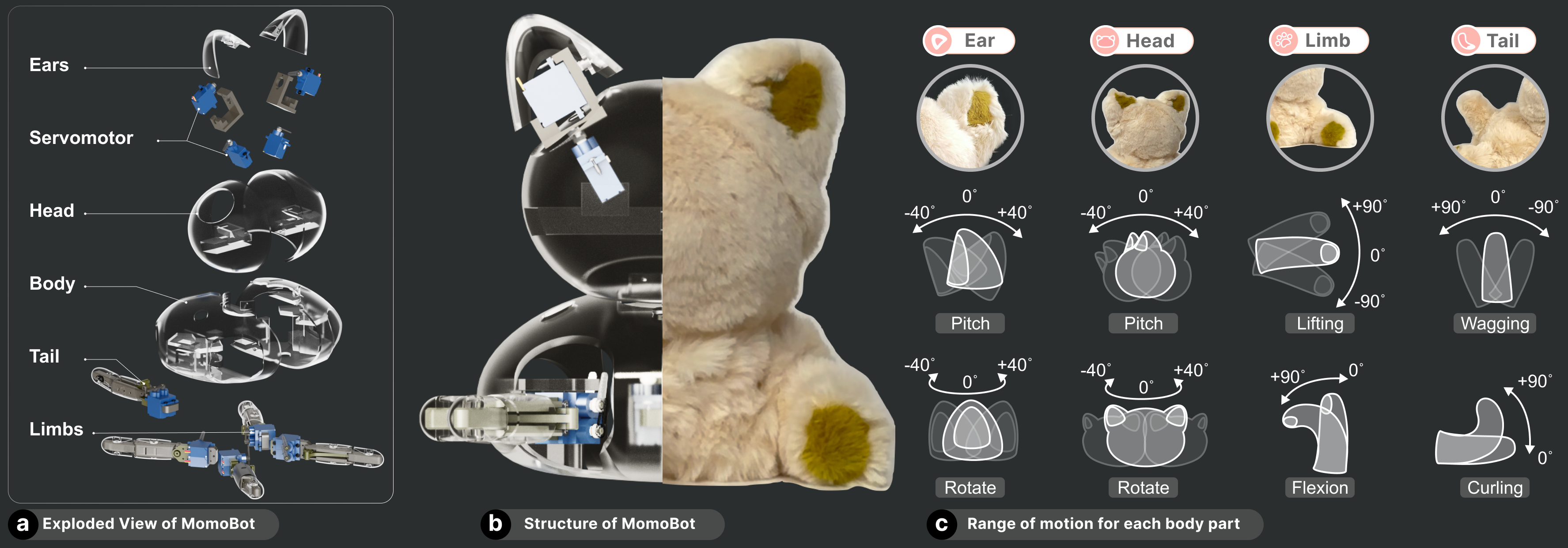}
    \caption{The structure of MomoBot. (a) Exploded view of MomoBot’s internal components with servomotors. (b) Structural cutaway showing the plush exterior and embedded mechanisms. (c) Motion ranges of ears, head, limbs, and tail across pitch, rotation, lifting, flexion, wagging, and curling.}
    \Description{A three-panel figure illustrating the design and motion capabilities of MomoBot, an animal-inspired robot. Left: Exploded view highlighting components including servomotors, head, ears, body, tail, and limbs. Center: Integrated internal and external structure. Right: Range of motion visualizations for ears (pitch/rotation), head (pitch/rotation), limbs (lifting/flexion), and tail (wagging/curling), with angular degrees annotated. }
    \label{fig:momobot}
\end{figure*}

Building on the reference cards, MojiKit is a design probe that supports users in exploring animal-inspired emotional robot behaviors through a workflow from knowledge inspiration to embodied prototyping and iteration. The system comprises three components (see Figure~\ref{fig:teaser}): (1) \textbf{Design Reference Cards}, offering structured inspiration from ethology and human–pet interaction analysis; (2) \textbf{Zoomorphic Robot Prototype (MomoBot)}, with articulated limbs, ears, and a tail for enacting animal-like movements; and (3) \textbf{Behavior Control Studio}, a parametric environment for authoring and refining behaviors. Importantly, MojiKit is not intended to prescribe executable solutions but to act as a design probe, stimulating creativity and encouraging novel interpretations beyond the prototype itself.

\subsection{Design Reference Cards from Interaction Analysis and Ethology}

The Design Reference Cards form the core resource of MojiKit, providing users with a modular and interpretable design space for creating affective behaviors of pet-inspired robots. Rather than prescribing a rigid sequence, the card set enables flexible entry points, encouraging exploration across multiple dimensions of human–pet interaction. A key feature of the cards is their dual perspective: they capture both human-centric and animal-centric viewpoints, supporting the design of robot behaviors that go beyond mimicry to foster meaningful relational bonds. Detailed card descriptions can be found in Section \ref{sec:reference cards}.

\subsection{Zoomorphic Robot Prototype (MomoBot)}

The hardware probe, named MomoBot, is a plush zoomorphic robot that embodies animal-like morphology while deliberately leaving its identity open for interpretation. Inspired by domestic cats and dogs, MomoBot has articulated ears, a head, four limbs, and a tail, yet no facial features (see Figure \ref{fig:momobot}). This design choice shifts attention away from facial expressions toward bodily motion, while also reducing constraints on users’ imagination.

MomoBot comprises eight articulated structures (two ears, head, four limbs, and tail), each with two degrees of freedom, for a total of sixteen actuated joints (see Figure \ref{fig:momobot}). The ears and head can rotate (±40°) and pitch (±40°), enabling turning, nodding, and ear folding/upright movements. The limbs support paw lifting (±90°) and flexion (0–90°), affording interactive gestures such as reaching or curling. The tail allows lateral wagging (±90°) and upward curling (0–90°), supporting both rhythmic and affective movements. The mechanical design of the limbs and tail builds upon the open-source project Amazing Hand by Pollen Robotics\footnote{\url{https://github.com/pollen-robotics/AmazingHand}}. Each structure employs a parallel mechanism actuated by two small servomotors to realize combined flexion/extension (curling) and lifting (wagging) motions. We adapted this principle into a zoomorphic morphology and integrated it with custom 3D-printed body parts, resulting in a cohesive robot body composed of eight moving structures and a fixed torso.

For actuation, sixteen servomotors (SG90) are driven by an Arduino Mega2560 controller with a custom-designed expansion board, while an ESP32 WiFi module enables wireless communication with the Behavior Control Studio. All structural components were fabricated using PLA 3D printing, assembled with standard screws, and arranged with internal slots to secure joints and electronic components. The prototype weighs approximately 3 kg (comparable to an adult house cat) and was built at relatively low cost (approximately~45 USD including servos, controllers, PLA material, and batteries).


\begin{figure*}
    \centering
    \includegraphics[width=\linewidth]{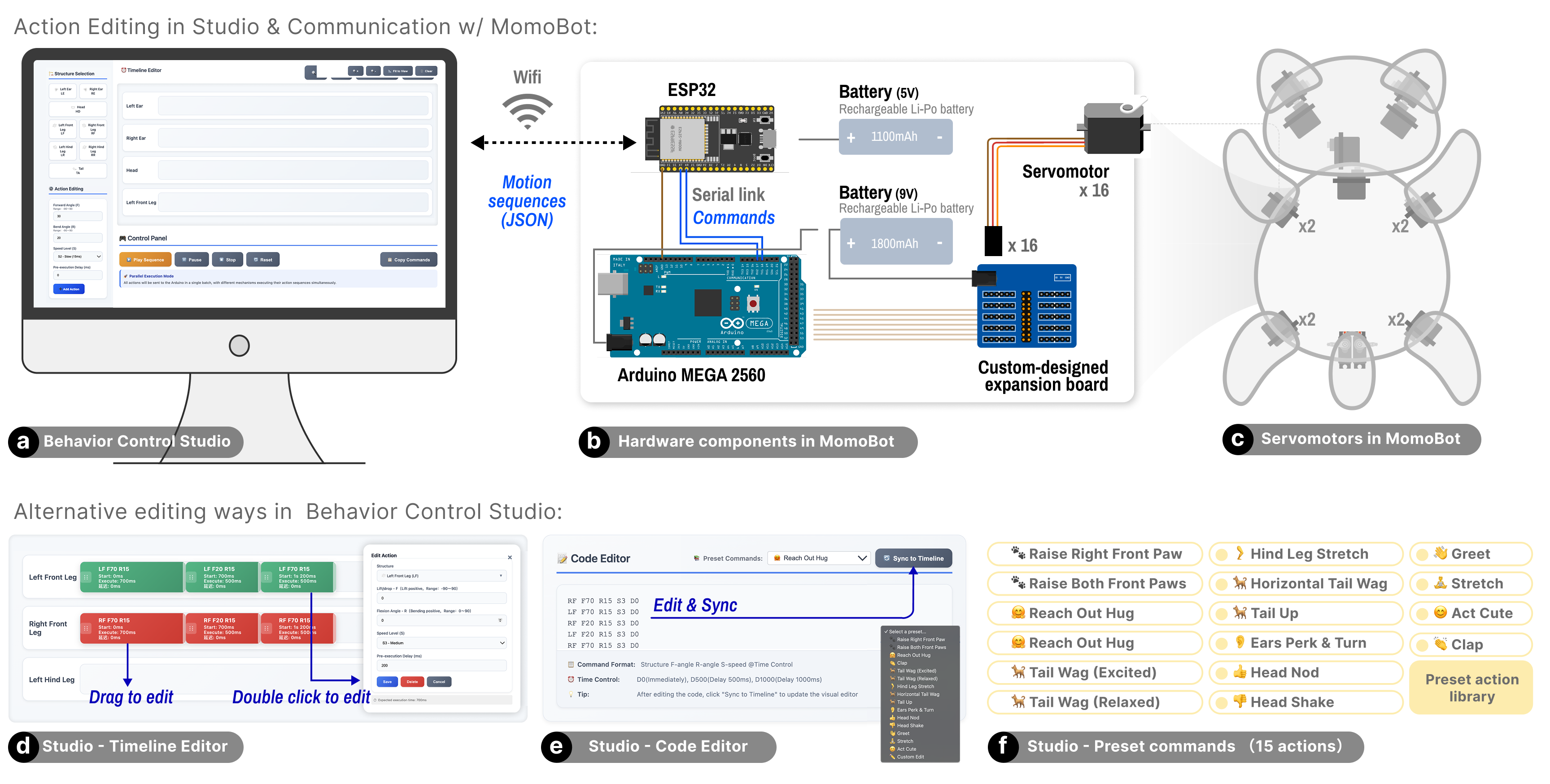}
    \caption{Action editing and execution in MomoBot. (a) Motion sequences created in the Behavior Control Studio are (b) sent via Wi-Fi to the ESP32 and relayed as serial commands to the Arduino Mega, (c) which drives 16 servomotors. The Studio supports (d) timeline editing, (e) code editing, and (f) preset actions.}
    \Description{A multi-panel figure describing the "Behavior Control Studio" for editing and communicating motion sequences to the MomoBot robot via WiFi. (a) Main studio interface for sequencing actions. (b) Hardware components: ESP32, Arduino MEGA 2560, batteries, and a customized expansion board. (c) Diagram of 16 servomotors distributed across MomoBot's body. The lower half of the diagram is Alternative editing methods: (d) Timeline editor with drag and double-click interactions, (e) Code editor with timeline synchronization, (f) Preset command library with 15 pre-built actions. }
    \label{fig:behavior control studio}
\end{figure*}

\subsection{Behavior Control Studio}

The Behavior Control Studio provides a software environment that enables users to author, preview, and refine expressive motion sequences on the zoomorphic prototype. To bridge high-level design activities with low-level actuation, the Studio adopts a three-layer architecture: a web-based interface, an ESP32 WiFi relay, and an Arduino Mega2560 controller. This architecture separates interaction design from hardware control, allowing users to focus on behavior creation while ensuring robust execution on the physical robot.

On the web-based interface, users interact with a timeline-based editor that allows drag-and-drop composition of motion blocks (Figure \ref{fig:behavior control studio} (d)). Each block specifies movement parameters—front-lift angle (F), rotational/bending angle (R), speed level (S), and execution delay (D)—and can be positioned and scaled in time. A built-in preview renders the robot’s body structures, highlighting active components during playback. To support rapid prototyping, the interface includes a preset action library (15 actions, e.g., paw lifting, nodding, tail wagging) and import/export functionality for sharing sequences (Figure \ref{fig:behavior control studio} (f)). Direct code editing is also available for advanced users (Figure \ref{fig:behavior control studio} (e)).

The PC connects to the robot through an ESP32 relay, which bridges the web interface and the Arduino Mega2560 via WiFi-to-serial communication (Figure \ref{fig:behavior control studio} (a,b,c)). This setup decouples interaction design from low-level hardware control while ensuring reliable command delivery. The ESP32 translates requests from the web interface into serial communication, forwarding them to the Arduino controller at 115200 baud. Lightweight error handling, including timeouts and retries, ensures stable performance in workshop and lab contexts.

At the Arduino control layer, sixteen servomotors corresponding to the robot’s limbs, ears, tail, and head are managed by a queue-based execution engine. A cubic Bézier smoothing algorithm interpolates trajectories between target positions, producing fluid and lifelike motion rather than abrupt changes. Multiple structures can execute independent sequences concurrently, enabling complex, coordinated full-body behaviors.


\subsection{User Flow} \label{sec:user flow}

Users start by drawing inspiration from the Design Reference Cards, then create motion sequences in the timeline-based editor of the Behavior Control Studio. The designed sequences are transmitted wirelessly via WiFi to MomoBot, where users can immediately observe the corresponding motions in real time. Based on these embodied effects, users reflect on the outcomes and refine their designs through iteration. To lower the entry barrier, the Studio also provides 15 preset cases (Figure~\ref{fig:behavior control studio} (f)) that can be directly tested, modified, or used as starting points for custom designs. This workflow positions MojiKit as a design probe that offers immediate embodied feedback and encourages creative exploration beyond the prototype’s built-in repertoire.

We used an LLM-based coding assistant (GPT-4o via Cursor IDE) to support the initial drafting of the web-based interface for the Behavior Control Studio. The generated code was subsequently reviewed, debugged, and adapted by the authors to ensure correctness and integration with the ESP32–Arduino architecture.

\section{Study Method: the Co-design Workshop}
\begin{table*}[]
\caption{Participant Demographics and Animal Interaction Experience.}
\Description{A table titled ``Participant Demographics and Animal Interaction Experience'' with 18 participants (P01-P18) across 9 groups (G1-G9). Columns list participant ID, gender (F/M), age, professional background (primarily design-related fields), pet ownership experience (years or N/A), and a summary of their animal interaction habits (e.g., daily care, occasional observation). }
\label{tab:workshop-demographics}
\begin{tabular}{@{}cccclll@{}}
\toprule
\multicolumn{2}{c}{\textbf{Group/ID}} & \textbf{Gen} & \textbf{Age} & \multicolumn{1}{c}{\textbf{Background}} & \multicolumn{1}{c}{\textbf{Pet Exp.}} & \multicolumn{1}{c}{\textbf{Animal Interaction Summary}} \\ \midrule
G1 & P01 & F & 26 & Interaction Design & N/A & Occasional park encounters; touch/observation \\
 & P02 & F & 27 & Interaction Design & Dog (7 yrs) & Daily care: feeding, grooming, play, health \\
G2 & P03 & F & 26 & \begin{tabular}[c]{@{}l@{}}Biomedical Eng. \\ \& Robotics Design\end{tabular} & Dog (3+ yrs) & Regular care: feeding, grooming, play \\
 & P04 & F & 25 & \begin{tabular}[c]{@{}l@{}}Urban \& Rural \\ Planning Design\end{tabular} & N/A & Occasional visits; play/observation \\
G3 & P05 & M & 31 & \begin{tabular}[c]{@{}l@{}}Information Art \\ \& Design\end{tabular} & Dog (5 yrs) & Deep care; frequent play; behavioral familiarity \\
 & P06 & F & 29 & Interaction Design & N/A & Frequent pet cafés/friends’ pets; casual interaction \\
G4 & P07 & M & 28 & \begin{tabular}[c]{@{}l@{}}Photoelectric \\ Sensor Design\end{tabular} & Cat (2 yrs) & Daily care: feeding, grooming, medical support \\
 & P08 & M & 27 & Engineering Mechanics & N/A & Occasional café visits; observation \\
G5 & P09 & F & 26 & \begin{tabular}[c]{@{}l@{}}Industrial \& Visual \\ Design\end{tabular} & Cat (3 yrs) & Daily care: feeding, play, grooming \\
 & P10 & F & 27 & Design & N/A & Occasional contact via friends; touch/play \\
G6 & P11 & M & 27 & Interaction Design & Cat (4 yrs) & Daily care and bonding; pet with family \\
 & P12 & F & 25 & HCI / Design & Cat (1.5 mo) & Early bonding and care tasks \\
G7 & P13 & F & 30 & Interaction Design & Cat (5 yrs) & Long-term care and interaction \\
 & P14 & F & 25 & Industrial Design & Cat (1.5 yrs) & Daily routines; behavioral observation \\
G8 & P15 & F & 27 & Industrial Design & N/A & Minimal contact; occasional park observation \\
 & P16 & F & 25 & \begin{tabular}[c]{@{}l@{}}Industrial \& Interaction \\ Design\end{tabular} & Dog (1 yr) & Limited experience; occasional care/play \\
G9 & P17 & F & 26 & Interaction Design & N/A & Occasional cafés/friends’ pets; limited interaction \\
 & P18 & F & 26 & Industrial Design & N/A & Mostly observation; little physical interaction \\ \bottomrule
\end{tabular}
\end{table*}

To investigate how users create emotionally expressive behaviors for animal-inspired robots using MojiKit, we conducted nine co-creation workshops. 

\subsection{Participants}

We recruited 18 participants (13 female, 5 male; $M_{\text{age}} = 26.83$, $SD = 1.67$) through social media and email announcements. All had foundational knowledge in interaction or robotic design. To examine the influence of animal-related experience, 10 participants reported pet ownership or frequent interaction with animals, while 8 reported no such experience. Groups were assigned to three conditions based on pet ownership status: experienced (both owners), inexperienced (neither), and mixed. Each session involved two participants and was facilitated by two researchers. Participants provided informed consent and received 15 USD compensation plus a 2 USD transportation subsidy. Table~\ref{tab:workshop-demographics} summarizes the demographics. All procedures were approved by the Institutional Review Board (IRB) of Hong Kong University of Science and Technology (Guangzhou) (HKUST(GZ)-HSP-2025-0186).





\begin{figure*}
    \centering
    \includegraphics[width=\linewidth]{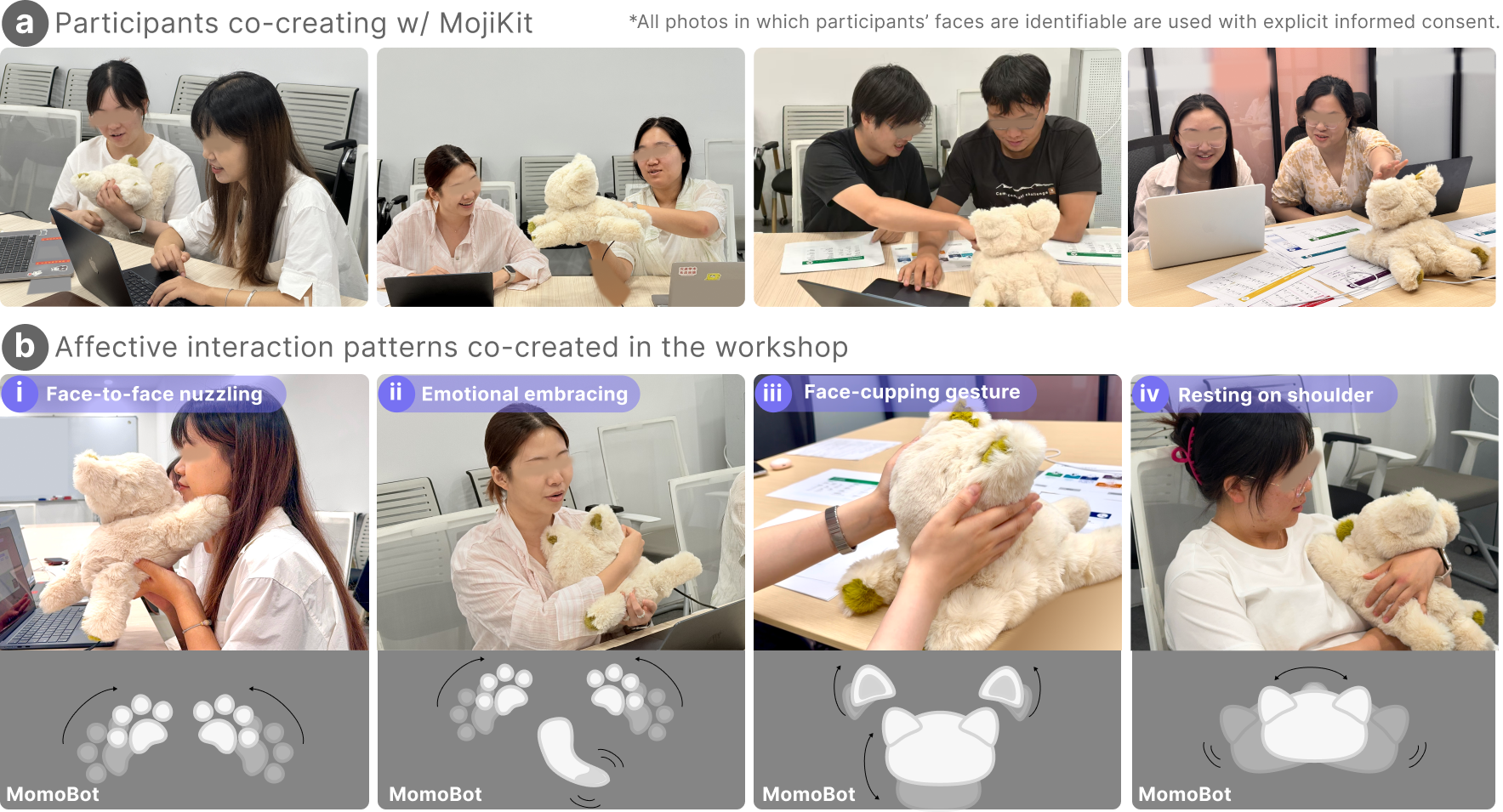}
    \caption{Co-design workshop.(a) Participants co-creating with MojiKit during the workshop sessions. (b) Examples of affective interaction patterns co-created by participants, including (i) face-to-face nuzzling, (ii) emotional embracing, (iii) face-cupping gesture, and (iv) resting on the shoulder. Photos are used with informed consent.}
    \Description{A two-panel figure from co-creation workshops. (a) Four example groups of participants collaboratively designing with MojiKit around a table, using plush toys in a well-lit indoor environment. (b) Four affective interaction pattern examples developed during the workshop: (i) face-to-face nuzzling, (ii) emotional embracing, (iii) face-cupping gesture, and (iv) resting on the shoulder, each shown with participant photos (faces blurred for privacy) and corresponding MomoBot robotic motion demonstrations. }
    \label{fig:workshop}
\end{figure*}

\subsection{Procedure}

Each workshop session lasted approximately 100 minutes and was structured into five consecutive stages. The process was designed to scaffold participants from orientation to hands-on design gradually, and finally to reflection, ensuring both creative exploration and systematic data collection:  

\begin{itemize}
  \item \textbf{Introduction \& consent (5 min).} Participants were welcomed and briefed on the study objectives and data privacy protocols. They signed the consent form and completed a background questionnaire about prior experience with pets and interaction design.  

  \item \textbf{Tutorial (10 min).} The researcher introduced the concept of MojiKit. A demonstration of the web-based interface (Behavior Control Studio) for real-time MomoBot control and the printed Design Reference Cards was provided. To illustrate the creative process, a complete example of designing a ``soothing'' interaction mode was presented.  

  \item \textbf{Design task (50 min).} Working in pairs, participants collaboratively designed 3–5 emotional interaction modes for MomoBot. Each mode included an intended emotion, a triggering condition (e.g., user behavior, environmental cues), a combination of robot behaviors (e.g., tail wagging, ear tilting), and parameter settings such as speed and amplitude. Participants could instantly test and refine their designs on MojiKit (see Section~\ref{sec:user flow}). Researchers passively observed and documented participants’ behaviors and verbal exchanges.  

  \item \textbf{Presentation (15 min).} Participants finalized and demonstrated their designs on the MojiKit robot. For each mode, they described its emotional intent, application scenario, and the rationale behind movement choices and parameter adjustments.  

  \item \textbf{Feedback and reflection (20 min).} Participants completed questionnaires on creativity support, system usability, and interaction experience. This was followed by a semi-structured interview on design strategies, toolkit usage, and reflections on limitations or areas for improvement.  
\end{itemize}

Participants were encouraged to explore, iterate, and refine their ideas through direct manipulation of MojiKit’s parameters (see Figure~\ref{fig:workshop}). The printed Design Reference Cards provided continuous reference for guidance, inspiration, and verification. To enable comprehensive analysis, all system interactions, design outcomes, and collaborative dialogues were audio–video recorded.

\subsection{Measurements}

We adopted a mixed-methods approach to capture both the design process and participants’ subjective evaluations. This allowed us to triangulate between quantitative metrics, self-reported experiences, and rich qualitative insights into collaborative behaviors.  

\subsubsection{Creativity Support Index (CSI)}  
The Creativity Support Index (CSI) \cite{cherry2014quantifying} was administered to evaluate participants’ perceptions of creative support across six dimensions (e.g., collaboration, exploration, expressiveness). CSI was chosen because it provides a standardized measure of how well interactive systems facilitate creativity, which is directly relevant to MojiKit’s intended role as a design probe.  

\subsubsection{Self-reported scales}  
In addition to CSI, participants completed two customized questionnaires: (1) a targeted evaluation of the Design Reference Cards (covering efficiency, inspiration, and clarity), and (2) a MomoBot \& Behavior Control Studio usability scale focusing on parameter adjustment, real-time feedback, and perceived intuitiveness. These scales provided structured insight into participants’ subjective experiences with the design resources, helping us assess whether the toolkit was usable, informative, and conducive to ideation. See Appendix \ref{sec:scales} for scale content. 

\subsubsection{Interviews \& video recording}  
To complement the quantitative data, we collected qualitative data through continuous video and audio recording, structured observation notes, and post-task semi-structured interviews. The interviews explored participants’ decision-making strategies, their use of the Design Reference Cards, and reflections on the co-creation experience. The recordings allowed us to capture nuanced behaviors such as gesture, turn-taking, and spontaneous remarks, which are essential for understanding how participants externalized and negotiated ideas. All verbal and visual data were transcribed, anonymized, and subjected to thematic analysis. See Appendix \ref{sec:interview} for interview framework. 

\subsection{Data Analysis}

We conducted both quantitative and qualitative analyses to evaluate the effectiveness of the design tools and understand participants’ design strategies.

Quantitative data from CSI and targeted questionnaires were analyzed using descriptive statistics to assess overall user experience. We also compared design outcomes (e.g., number and diversity of interaction modes, parameter variations) between participants with and without prior pet experience.

Qualitative data (interviews, worksheets, observation notes) were analyzed using thematic analysis \cite{terry2017thematic}. Two researchers independently conducted thematic analysis \cite{fereday2006demonstrating}, guided by predefined categories (e.g., design motivation, decision strategies, use of the Design Reference Cards), and refined the scheme through iterative discussion. Recurrent themes were identified around creativity, tool use, and emotional design reasoning. Co-created interaction patterns were then reviewed and categorized by emotional intent, triggering logic, and motion features to support future reuse in robot behavior evaluation and user studies.


\section{Results \& Findings}
\begin{figure*}
    \centering
    \includegraphics[width=\linewidth]{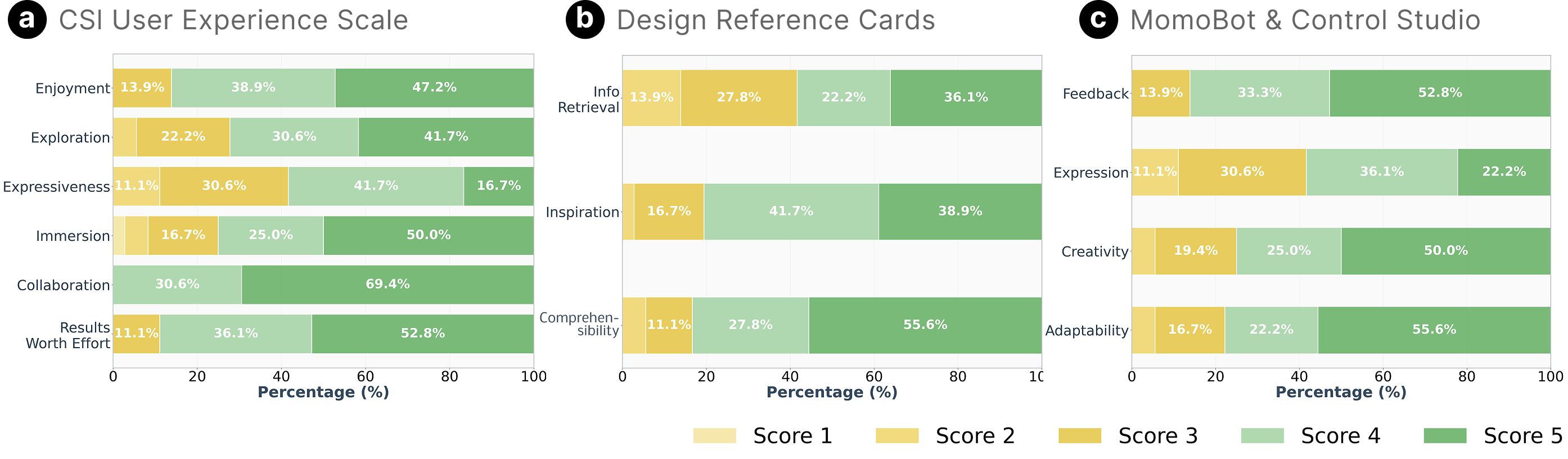}
    \caption{Quantitative evaluation results of MojiKit. (a) Creativity Support Index (CSI) \cite{cherry2014quantifying} assessing six dimensions of creative support, (b) Design Reference Cards evaluation focusing on efficiency, inspiration, and clarity, and (c) MomoBot \& Behavior Control Studio usability scale covering feedback comprehension, expressiveness, creative exploration, and adaptability.}
    \Description{Three horizontal stacked bar charts comparing quantitative evaluation results across three HCI systems. (a) CSI User Experience Scale: Six metrics (Enjoyment, Exploration, Expressiveness, Immersion, Collaboration, and Results Worth Effort) show score distributions (1-5), with Collaboration having the highest 5-score proportion (69.4\%, dark green). (b) Design Reference Cards: Three metrics (Information Retrieval, Inspiration, Comprehensibility), where Comprehensibility achieves the highest 5-score rate (55.6\%). (c) MomoBot & Control Studio: Four metrics (Feedback, Expression, Creativity, Adaptability), with Adaptability showing 55.6\% at 5-score. All charts use a green-to-yellow color gradient for scores (1=light yellow, 5=dark green) with percentage labels. }
    \label{fig:quantitative_results}
\end{figure*}

We report both quantitative evaluations and qualitative design outcomes. Scale-based measures assessed creative support, card usability, and prototype interaction, while workshop proposals revealed how participants envisioned companion interactions, extending beyond MojiKit’s built-in capabilities and highlighting insights into realism, identity, proactivity, and relationship styles. \Revision{As all result in FIgure \ref{fig:quantitative_results} use abbreviated keywords instead of full item texts, detailed descriptions of each scale and interview prompt can be found in Appendix~\ref{Evaluation Scales and Interview Protocol in workshop}.}

\subsection{Quantitative Results from Scales}

\subsubsection{Creativity Support Index (CSI)}

Participants’ creative experience was assessed using the Creativity Support Index (CSI) \cite{cherry2014quantifying}. Results indicated a generally positive experience. \textbf{Collaboration} scored highest ($M=4.69$, $SD=0.46$), followed by \textbf{Results Worth Effort} and \textbf{Enjoyment} (both $M>4.3$). The lowest score was for \textbf{Expressiveness} ($M=3.64$, $SD=0.89$), while \textbf{Immersion} showed the most variability ($SD=1.06$). These findings suggest strong collaborative and enjoyable experiences, with room to improve expressiveness and engagement.

\subsubsection{Evaluation of Design Reference Cards}

Evaluation of the card-based Design Space focused on \textbf{Comprehensibility}, \textbf{Inspiration}, and \textbf{Information Retrieval}. Participants rated the cards positively overall. \textbf{Comprehensibility} was highest ($M=4.33$, $SD=0.88$), with terminology and relationships between card types considered clear. \textbf{Inspiration} also scored well ($M\approx4.2$), while \textbf{Information Retrieval} was more variable ($M=3.56$–$4.06$). Overall, the cards supported ideation and understanding, though search efficiency could be improved.

\subsubsection{Evaluation of MomoBot}

MomoBot was evaluated across \textbf{Feedback Comprehension}, \textit{Expressive Capability}, \textbf{Creative Exploration}, and \textit{Design Adaptability}. The highest ratings were for \textbf{Feedback Comprehension} ($M>4.3$), showing real-time responses effectively clarified parameter–behavior links. \textbf{Creative Exploration} was also strong ($M\approx4.2$), while \textbf{Expressive Capability} was lower ($M \in [3.56, 4.06]$), indicating limited support for nuanced expression. \textbf{Design Adaptability} revealed that although not all intentions could be realized, participants adapted through parameter adjustments ($M\approx4.2$). Overall, MomoBot was valued for clarity and exploration, with opportunities to enhance expressiveness and flexibility.

\subsection{Designed Interaction Patterns Emerged from Workshops}

Using MojiKit, nine groups generated about 40 proposals; after consolidation, 35 valid interaction patterns were retained, each documented with intent, trigger, behavior, meaning, and interpretation (see Appendix \ref{sec:design outcomes}). Analysis revealed four themes: (1) companionship-oriented intents, (2) balance of human-initiated and proactive triggers, (3) expressive sequences from simple primitives, and (4) affective resonance through reciprocal appraisal.

Table~\ref{tab:stats_patterns} summarizes the 35 validated patterns. The most frequent human intents were greeting/reunion and affection/comfort (20.0\% each), followed by playfulness/teasing (17.1\%) and attention-seeking (14.3\%). Most interactions were triggered by human actions (60.0\%), though some relied on routines, environmental cues, or proactive robot behaviors. Common robot primitives included head turns, approaching, and tail wags (together 51.4\%). Overall, 71.4\% conveyed positive affect, underscoring participants’ expectation for robots to provide companionship and emotional support.

\subsubsection{Human Intents: Companionship at the Core}
Greeting and affection dominated the designs (e.g., G6-3 ``Welcome home'', G1-3 ``Pet me''), reinforcing the robot’s role as a social companion. Playful cases (e.g., G3-2 ``Playfulness'') highlighted humor, while training (e.g., G6-1 ``Sit training'') and corrective episodes (e.g., G4-1 ``Unexpected surprise'') reflected everyday care routines.

\subsubsection{Triggers: Predominantly Human-initiated, with Emerging Proactivity}
Most patterns were triggered by human actions such as calling (e.g., G1-1 ``Who is calling me''), petting (e.g., G1-2 ``Don’t touch me''), or feeding (e.g., G1-4 ``A gift from mom''). Others aligned with routines (e.g., G6-2 ``Good morning, good night'') or proactive robot initiatives, such as G7-1 ``I want attention'', where the robot banged its limbs to demand engagement.

\subsubsection{Robot Behavior: Expressive Sequences from Simple Primitives}
Designs drew on a small set of ethologically inspired primitives (head-turning, approaching, tail-wagging, vocalization, paw-tapping, and lying down) combined into expressive sequences. These combinations illustrate how simple primitives can generate nuanced affective meanings:

\begin{itemize}
    \item Greeting: approach + wagging + vocalization (e.g., G5-1B ``Hi!!!'').
    \item Affection: lying down + curling + purring (e.g., G8-1 ``Lazy cat'').
    \item Avoidance: head-turn + retreat (e.g., G1-2 ``Don’t touch me'').
\end{itemize}

At the same time, many final proposals exceeded MomoBot’s built-in capabilities, showing that the tool encouraged imagination beyond its own scope. Examples included \textbf{behavioral capabilities} (e.g., P10’s design of a robot greeting its owner at the door, P12’s design reminding her to sleep), \textbf{sound capabilities} (e.g., P17 and P18 incorporating purring sounds to enhance intimacy), and even \textbf{thermal expression} (e.g., P4 envisioning a robot that warms up during naps, P6 imagining one that rests on the user’s stomach to provide warmth and companionship). These extended ideas highlight MojiKit’s role as a design probe, stimulating multimodal concepts unconstrained by the current prototype.

\subsubsection{Affective Meaning and Human Interpretation}
Over 70\% of designs expressed positive emotions such as care, joy, and intimacy (e.g., G9-1 ``I miss you'', G6-2 ``Good morning, good night''). Negative or corrective behaviors (e.g., G4-1 ``Unexpected surprise'') introduced boundaries. Participants also annotated their own feelings, such as being cared for, amused, or compelled to comfort, revealing a reciprocal framing of human–robot interaction.

\subsubsection{Design Sources and Strategies: Pet Experience Anchors Designs, Cards Enable Creative Hybridization} \label{sec:design source}
Participants’ designs drew on varied sources, including personal pet-keeping experience (e.g., G1-1 ``Who is calling me''), imagination based on animation or puppetry (e.g., G4-2 ``Hello world/stranger''), and MojiKit reference cards (e.g., G6-2 ``Good morning, good night''). Scenarios spanned daily routines, homecoming rituals, and leisure play, situating MomoBot in intimate everyday contexts.

In sum, the 35 patterns show how participants used \textit{MojiKit} to create a wide repertoire of companion scenarios, extending beyond MomoBot’s motor capabilities to multimodal interactions of movement, sound, and routines. The toolkit thus functioned as a design probe rather than merely a prototype. Design sources varied: most ideas came from personal pet experience (18), with others from imagination (5), reference cards (3), or mixed sources (9). This diversity illustrates how structured resources can scaffold creativity while leaving room for personal experience and imagination.

\begin{table*}[]
\caption{Summary statistics of 35 validated interaction patterns collected from the workshop.}
\Description{A four-section statistical table summarizing 35 validated human-robot interaction patterns from co-design workshops. Section 1 (Human Intent): Eight categories including Greeting/Reunion (20.0\%), Affection/Comfort (20.0\%), and Play/Teasing (17.1\%) with behavioral examples. Section 2 (Trigger Types): Four trigger categories dominated by Human Action (60.0\%) including speech/touch/gesture. Section 3 (Robot Behavior): Eight response types including Head-turn/Nodding (20.0\%), Approach/Move closer (17.1\%), and Tail-wagging (14.3\%). Section 4 (Affective Meaning): Five affect categories with Positive-Seeking behaviors (57.1\%) as most prevalent. }
\label{tab:stats_patterns}
\begin{tabular}{@{}lccl@{}}
\toprule
\textbf{Category} & \textbf{Count (n)} & \textbf{Percentage (\%)} & \textbf{Examples} \\ \midrule
\multicolumn{4}{l}{\textbf{Human Intent}} \\ \midrule
Greeting / Reunion & 7 & 20.0\% & Running to greet; playful salutation \\
Affection / Comfort & 7 & 20.0\% & Sleeping together; cuddling; soothing \\
Play / Teasing & 6 & 17.1\% & Toy play; playful provocation \\
Attention-seeking & 5 & 14.3\% & Responding to name; seeking attention \\
Training / Instruction & 3 & 8.6\% & Sit command; learning tricks \\
Boundary / Discipline & 3 & 8.6\% & Correction after misbehavior \\
Avoid / Refuse & 2 & 5.7\% & Head-turning away; resisting touch \\
Other / Unclear & 2 & 5.7\% & Incomplete or ambiguous designs \\ \midrule
\multicolumn{4}{l}{\textbf{Trigger Types}} \\ \midrule
Human Action & 21 & 60.0\% & Calling name; stroking; feeding \\
Environmental Cue & 5 & 14.3\% & Sounds; objects; unexpected events \\
Temporal Routine & 5 & 14.3\% & Wake-up; bedtime; returning home \\
Proactive Robot & 4 & 11.4\% & Robot initiates when ignored \\ \midrule
\multicolumn{4}{l}{\textbf{Robot Behavior}} \\ \midrule
Head-turn / Nodding & 7 & 20.0\% & Turning head toward sound \\
Approach / Move closer & 6 & 17.1\% & Walking toward the human \\
Tail-wagging & 5 & 14.3\% & Rapid wagging when excited \\
Paw-tapping / Contact & 4 & 11.4\% & Patting; pawing; leaning \\
Vocalization & 3 & 8.6\% & Bark; meow; purring sound \\
Lie-down / Curl / Roll & 3 & 8.6\% & Rolling over; curling up \\
Retreat / Avoidance & 3 & 8.6\% & Withdrawing; avoiding touch \\
Other complex sequences & 4 & 11.4\% & Running around; chasing; jumping \\ \midrule
\multicolumn{4}{l}{\textbf{Affective Meaning of Robot Behavior}} \\ \midrule
Positive--Seeking & 20 & 57.1\% & Approaching; cuddling; wagging \\
Positive--Comforting & 5 & 14.3\% & Lying nearby; curling beside human \\
Negative--Avoiding & 3 & 8.6\% & Turning away; leaving \\
Disciplinary / Corrective & 2 & 5.7\% & Stiff posture after scolding \\
Ambiguous / Mixed & 5 & 14.3\% & Multiple or unclear affect \\ \bottomrule
\end{tabular}
\end{table*}

\subsection{Empowering the Design Process with MojiKit}

From interviews and observations, we examined how MojiKit supported participants’ design process. By combining structured resources (Reference Cards) with an operable prototype (MomoBot), it lowered creative barriers, offered conceptual scaffolding, and inspired ideas beyond the tool’s built-in scope. 

\subsubsection{Structured \& Flexible Design Pathways}

The Design Reference Cards offered a structured framework by breaking human–pet interactions into distinct modules. This modular approach helped participants identify key components and start ideation from any entry point. For example, P7 noted that the cards guided her to start from the human perspective, which made ideation easier despite her limited knowledge of pet behavior. Others (P8, P13) highlighted that the structure gave them confidence to explore ideas without missing steps. P12 noted that the cards helped transform fragmented ideas into coherent designs. P5 highlighted their value in providing a traceable path for prototyping and iteration.

With this structured support, participants developed three typical design pathways:

\begin{itemize}
    \item \textbf{Behavior-First Pathway:} Starting from a robot action and tracing back to its emotional meaning and context, often during prototype debugging (e.g., observing Momo’s motion as ``drawing close'').
    \item \textbf{Scenario-First Pathway:} Beginning with a life scenario or emotional state, identifying human needs, and then defining robot behaviors (e.g., designing gentle patting in a ``comforting'' scenario).
    \item \textbf{Re-enactment Pathway:} Reproducing a full human–pet interaction by mapping human states, environmental cues, and animal behaviors (e.g., recreating a ``welcoming home'' scene).
\end{itemize}

Overall, Reference Cards strike a balance between structure and flexibility. The cards provide a clear scaffold that ensures interaction patterns are complete, while preserving exploratory space for participants with diverse backgrounds. This allows users to combine personal experiences, imagination, and prototype experimentation into varied design pathways.

\subsubsection{Lowering the Creative Barrier}

The code-free nature of MojiKit lowered the creative barrier, enabling participants without technical backgrounds to engage directly in design through cards and hardware. Predefined examples provided multiple entry points, supporting exploration, validation, and iteration. Several participants highlighted the value of starting with examples. P1 described the Studio as ``very simple and quick to learn,'' while P17 preferred adjusting templates before adding new ideas. P9 noted that modifying an existing action gave her a concrete starting point, and P10 emphasized that examples allowed immediate testing of whether behaviors conveyed the intended emotion. Others used examples for combining preset actions (e.g., P14), drawing inspiration from real-time interactions (e.g., P17), or validating feasibility (e.g., P6). Together, these findings show that MojiKit reduced the learning curve and maintained engagement, allowing participants to ``play while designing'' (P1) and gradually build confidence through iterative experimentation.

\subsubsection{MojiKit Stimulated Creativity Beyond Its Built-in Scope}

MojiKit reduced barriers and inspired participants to expand ideas beyond the tool’s inherent limitations as probe. Inspiration emerged at two levels: the \textbf{cards}, which guided reflection and extension of the design space, and the \textbf{prototype (MomoBot)}, which evoked new emotional associations through embodied interaction.

At the \textbf{card level}, participants drew insights on both macro and micro scales. Macroscopically, the ``environmental elements'' module prompted reflection on indirect emotional effects. P3 noted that it highlighted the observer’s role, leading him to consider how robot–environment interactions can shape human feelings. Microscopically, human- and animal-centric items encouraged deeper analysis of motivations and needs. For example, P8 linked ``praising animals'' to training scenarios, and P15 realized that quietly watching independent actions was her preferred interaction type. Participants also went beyond the cards, suggesting additions such as delivery boxes, toys, or lighting changes (P3), illustrating how the cards triggered active extension of the design space. At the \textbf{prototype level}, interacting with MomoBot further inspired new ideas. Participants often reinterpreted predefined motions as affective expressions. For example, P17 perceived ``Raise Both Front Paws'' as begging for attention and designed around it, while P14 humorously described a debugging motion as a ``tantrum'' and built on that theme. P13 associated table-tapping with animal-like frustration. Direct touch interactions (see Figure~\ref{fig:workshop}b) also encouraged participants to explore emotional scenarios physically. These cases show that the hardware stimulated, rather than constrained, creative exploration.


\subsection{Design Insights: Realism, Identity, Proactivity, and Relationship Styles in Pet Robot Interaction}

Beyond specific interaction patterns, participants’ designs revealed broader insights into how they envision pet robot companionship. These insights clustered into four trajectories: \textbf{behavioral realism}, the balance between mimicking real animals and prioritizing emotional needs; \Revision{\textbf{morphological abstraction}, the tension between open imaginative interpretation and the need for emotionally legible cues;} \textbf{robot identity}, expectations for a ``perfect pet'' versus an ``imperfect companion''; \textbf{interaction proactivity}, preferences for strong versus subtle modes of engagement; and \textbf{relationship styles}, the extent of responsibility and reciprocity participants were willing to accept in long-term companionship.

\subsubsection{Behavioral Realism vs.\ Emotional Needs in Pet Robot Design}
Participants showed two orientations in their design: some pursued realism, aiming to make the robot closely resemble a real pet, while others emphasized emotional connection, extending beyond real animal behaviors to prioritize human needs. For example, P9 initially likened the robot to a cat or dog due to its ears and limbs, but after observing its motions, she concluded ``it is not an animal'' and shifted to assigning more human-like actions to foster emotional connection. In contrast, P10 argued that ``being more like a real animal makes it easier for me to invest emotionally,'' highlighting expressions and details as critical for bonding. This divide was also shaped by pet experience: participants with pet-keeping backgrounds tended to avoid abilities beyond real animals’ capabilities (e.g., P16 noting that ``cats don’t comfort people''), whereas those without pets (e.g., P15, P12) expected the robot to ``proactively comfort'' them when sad.

\Revision{
\subsubsection{Abstract Morphology: Open Imagination vs. Emotional Legibility} \label{Abstract Morphology}
Participants also reflected on MomoBot’s abstract morphology, such as its lack of facial features. Overall, 16 out of 18 participants perceived the featureless form as a strategy that lowers expectations and shifts attention toward motion-based cues rather than species-specific facial conventions. P9 noted that although MomoBot did not display facial expressions, the presence of a discernible ``head'' offered sufficient directional cues to signal ``it is interacting with me.'' Similarly, P12 described the face-free design as making the robot feel more like an undefined ``companion creature,'' opening greater imaginative space for interpreting its behaviors. P15 further noted that she did not feel the need for facial expressions; adding a face would raise expectations for human-like expressiveness and risk pushing the robot toward the uncanny valley.

At the same time, some participants (e.g., P10, P16) pointed out that without eyes or facial expressions, certain emotional states were harder to infer directly from appearance and required more pronounced motion or postural changes to be legible. This observation aligns with the relatively lower ratings for Expressiveness in the quantitative evaluation, suggesting that while abstraction foregrounds motion as the primary expressive channel, it also constrains the robot’s ability to convey fine-grained emotional nuances.
}

\subsubsection{Robot Identity: Between ``Perfect Pet'' and ``Imperfect Companion''} \label{Robot Identity}
Participants projected both ideals and tolerances when imagining robotic companionship. Some accepted limitations, drawing parallels with their own pets: P2 expressed being ``more tolerant'' of the robot’s clumsy mobility because ``my dog isn’t that bright anyway,'' even reflecting that she was ``too demanding of her dog'' and hoped the robot could compensate. In contrast, others expected the robot to serve as an idealized companion. P1 emphasized a desire for the robot to ``perfectly satisfy emotional needs'' without clumsy behaviors, while P12 hoped it could ``love me one hundred percent'' and provide absolute emotional security. These perspectives show that while some users value pets’ imperfections as authentic traits, others expect robots to transcend them and become perfected emotional proxies.

\subsubsection{Interaction Proactivity: Strong vs.\ Subtle Preferences}
Design outcomes also reflected differing preferences for interaction intensity. Some participants desired proactive and direct engagement: P16 wanted the robot to greet her at the door and show affection upon approach, while P12 stressed that only direct contact created a ``definite sense of closeness.'' Others preferred subtle presence: P15 noted that she did not need frequent initiations, as the robot’s independent movement around the house already provided companionship. These differences highlight varied expectations for interaction intensity, tied to individual personalities and distinct understandings of companionship.

\subsubsection{Relationship Styles with Companion Robots} \label{Relationship Styles}

Participants articulated three distinct styles in imagining their relationship with zoomorphic companion robots.  
\begin{itemize}
    \item \textbf{Responsibility-Exempt:} Robots were viewed as “one-way companionship proxies,” providing emotional value without requiring caregiving duties. This style emphasized comfort and presence while rejecting obligation (e.g., P1).  
    \item \textbf{Reciprocal-Caring:} Other participants imagined a bidirectional bond, where the robot would occasionally ``need'' them (e.g., feeding batteries), creating a sense of being responsible or useful. At the same time, they acknowledged that such expectations could also become burdensome, as in P15’s comment that being “waited for at the door” would feel overwhelming.  
    \item \textbf{Controlled-Boundary:} A third group framed responsibility as negotiable and switchable. They accepted caregiving moments as part of interaction, but defined them as contextual scripts rather than continuous obligations. This style emphasized user-defined boundaries to prevent long-term moral burden.  
\end{itemize}

Taken together, these styles illustrate that zoomorphic robots can elicit relational dynamics similar to those with pets, but that users actively negotiate how much responsibility they are willing to accept. This negotiation reflects broader tensions in HRI: between viewing robots as safe, responsibility-free companions, and attributing them needs that trigger reciprocity and obligation.

\section{Discussion}
This work introduced \textbf{MojiKit}, a design probe that combines structured data-informed resources with a zoomorphic prototype to support the creation of affective robot behaviors. Our study shows that MojiKit (1) scaffolds design processes and lowers creative barriers, (2) enables flexible exploration of emotional interactions across diverse user backgrounds, and (3) inspires ideas beyond the system’s current functions. These results address our research questions and set the stage for broader discussions of human–robot emotional bonds, ethical reflections, and implications for future zoomorphic companion robots.

\subsection{Resources and Tools for Supporting Animal-Inspired Robot Design}

Our study highlights the importance of providing designers with both structured reference resources and accessible development tools when creating emotionally expressive behaviors for animal-inspired robots, \Revision{while situating interactive robot motion in relation to traditional animation principles}.

\subsubsection{Structured Reference Cards for Affective Interaction Design} 
Designing emotionally expressive behaviors for animal-inspired robots is challenging because the process is often intuition-driven and idiosyncratic, making outcomes difficult to generalize or reuse \cite{obrist2013talking}. Existing HRI references highlight the importance of affective expression \cite{breazeal2003emotion}, but are largely descriptive, dispersed across literature \cite{fong2003survey}, or tailored to specific tasks and platforms \cite{kirby2010affective}, offering limited transferable guidance. Building on evidence that card-based toolkits effectively support ideation and communication in design \cite{roy2019card,peters2021toolkits,wolfel2013method}, MojiKit adopts a card format but contributes a systematically curated, domain-specific resource. By integrating ethological literature with coded human–pet observations, the cards articulate actionable components—human intents, animal behaviors and affect, context, triggers, and motion primitives. This structure lowered entry barriers, prompted consideration of overlooked aspects, and enabled both novices and experts to heuristically explore the emotional potential of zoomorphic companion robots.

\subsubsection{Code-free Platform with Real-time Feedback.} 

Designing robot behaviors is traditionally constrained by technical barriers: many existing development tools in HRI (e.g., Choregraphe for NAO \cite{hellou2023development}, visual programming language (VPL) interfaces \cite{vazquez2021visual}, or block-based environments such as Blockly \cite{fraser2015ten}) require programming knowledge, abstract representations, or rely on simulated execution \cite{dragule2021survey}. Such barriers limit participation from non-experts and make it difficult to iteratively test the emotional impact of embodied behaviors (\emph{need for accessible tools}). Responding to this gap, MojiKit provides a \emph{code-free, parameter-based interface} that directly controls a tangible zoomorphic prototype, delivering immediate embodied feedback. This design enables participants to observe how abstract emotional concepts translate into concrete motions, adjust parameters in real time, and refine ideas through rapid prototyping loops. By removing the requirement for programming and embedding real-time, physical feedback, the platform fosters creativity, lowers the entry threshold, and supports experimentation in ways not fully addressed by prior robot design environments.

\Revision{
\subsubsection{Connecting Interaction Strategies with Animation Principles}

During co-creation, some participants referenced motion patterns from animation, pointing to the rich body of design knowledge in animation and puppetry (Section \ref{sec:design source}). Classic principles, such as exaggeration and anticipation, are known to enhance the clarity of actions \cite{lasseter1998principles,johnston1981illusion}. MojiKit shares a central objective with these principles: prioritizing \textit{human interpretation} in design. While animation aims to ensure audiences correctly interpret a character's internal state \cite{wells2013understanding,johnston1981illusion}, MojiKit supports users in crafting behaviors that are legible to human observers. The two approaches can address distinct design dimensions: MojiKit assists users in determining specific interaction strategies, including proximity, touch, attention cues, and rhythm \cite{breazeal2003emotion,urakami2023nonverbal}; in parallel, animation principles primarily guide character setting and stylistic consistency \cite{reeves1996media,bates1994role}. Integrating MojiKit’s interaction logic with the stylistic frameworks of animation thus presents a valuable direction for future work, ensuring robot behaviors are both interactionally responsive and characteristically consistent.

}

\subsection{Design Implications}


From our findings, we derive \Revision{four} implications for designing animal-inspired companion robots: (1) supporting \textit{growth and adaptivity} to sustain long-term engagement, (2) ensuring \textit{perceived agency} as a foundation for emotional attachment, \Revision{(3) matching \textit{morphological cues}, including the use or avoidance of facial expression, to the robot’s actual expressive capabilities, }and (4) extending animal-like qualities into everyday contexts through \textit{distributed morphological augmentation}. Together, these implications view companion robots not as static artifacts but as evolving, agentive, and ecologically embedded presences in daily life.

\subsubsection{\textbf{Implication 1 – Designing for Growth and Adaptivity}}  

Participants emphasized growth as key to sustaining long-term engagement with animal-inspired companion robots. Beyond functional expansion, growth was framed as the accumulation of shared experiences and evolving relationships, countering the fading of novelty in social robots \cite{bartneck2024human,leite2013influence}. Three dimensions stood out: (1) gradual exploration and memory formation that personalize interactions \cite{fink2013living,luria2020destruction}; (2) relational trajectories moving from unfamiliarity to trust, echoing human–robot attachment accounts \cite{de2016ethical,weiss2009love}; and (3) new skills and more complex behaviors as markers of maturity, sustaining engagement through perceived learning \cite{kory2019long}. Growth was not always imagined as linear—participants also valued variation in habits, moods, or interaction styles to keep interactions fresh \cite{fernaeus2010you}. Overall, growth was seen as a mechanism for extending engagement and building emotional bonds, enabling users to feel they are ``living through time together'' with the robot \cite{reis2000handbook}.

\textbf{\textit{Design implication: Foster multidimensional ``growth'' by designing the robot’s dynamic relationships with environment and users, alongside its own learning and development.}} This entails (1) enabling robots to accumulate and surface memories that personalize interactions, (2) scaffolding relational trajectories that signal increasing familiarity and trust, and (3) incorporating both vertical progress (new skills, capabilities) and horizontal variation (moods, habits) to balance continuity with novelty. Such strategies can help transform companion robots from static artifacts into evolving partners, thereby sustaining engagement over time.

\subsubsection{\textbf{Implication 2 – Designing for Perceived Agency}}  

Participants emphasized robot agency as essential for emotional attachment. Agency was framed not as mere reactivity, but as situated autonomous presence—learning and adapting to users and environments (P12, P15), engage in self-initiated activities without constant human input, proactively initiate interactions, and to maintain routines, quirks, or idiosyncratic behaviors that convey individuality.. Notably, occasional non-critical ``bad behaviors'' (e.g., stubbornness or minor mistakes) were seen as enhancing authenticity and companionship. This aligns with findings that autonomy and limited unpredictability foster social attribution and engagement \cite{lemaignan2014dynamics,leite2013social,paepcke2010judging}, while excessive autonomy may threaten user comfort and control \cite{belpaeme2018social,duffy2003anthropomorphism}.  

\textbf{\textit{Design implication: Convey robot agency by designing firm preferences, non-interactive behaviors, and even subtle oppositional behaviors, thereby enhancing its role as a companion. }}This involves (1) embedding adaptive and self-initiated behaviors that signal independence, (2) creating routines, quirks, and even occasional non-critical ``bad behaviors'' that lend authenticity and personality, and (3) balancing autonomy with predictability to ensure user comfort. By calibrating this balance, designers can foster perceptions of robots as active partners rather than passive tools, thereby strengthening emotional attachment and sustaining long-term engagement.

\Revision{
\subsubsection{\textbf{Implication 3 – Rethinking Facial Expression in Zoomorphic Companions}}
Participants' divergent reactions to MomoBot’s faceless design, ranging from confusion to appreciation, highlight a critical design trade-off (Section \ref{Abstract Morphology}). On one hand, facial cues are often indispensable for readability, especially for robots intended to sustain complex social dialogues. Prior HRI research emphasizes that eyes and mouths are essential for directing attention and conveying discrete emotions (e.g., joy vs. surprise) with high immediacy and specificity \cite{breazeal2003emotion,fong2003survey}. On the other hand, our findings suggest that abstraction offers distinct advantages by shifting the user's focus toward motion-based cues (kinesics) \cite{duffy2003anthropomorphism,waytz2014mind} and mitigating the risk of the ``uncanny valley'' associated with realistic features \cite{blut2021understanding,paauwe2015designing}. Moreover, detailed faces often imply a level of social intelligence that simple robots cannot support, creating an \textit{expectation gap} \cite{paauwe2015designing}. Featureless design calibrates these expectations to the system’s actual limits, avoiding the frustration of unmet social promises.

\textbf{\textit{Design implication. Match morphological cues to actual emotional capabilities.}} Rather than prioritizing realism, designers should select morphological cues (including whether to use facial expression at all) that align with the robot’s intended expressive range and avoid implying capabilities the system cannot support.
}


\subsubsection{\textbf{Implication 4 – Morphological Extensions of Animal-like Qualities in Everyday Contexts}}  

Several participants moved beyond the paradigm of a single companion robot to imagine animal-like features embedded in everyday environments and artifacts. Examples included phones that “perk their ears” for reminders, cups with “tails” that prompt hydration, or beds resembling the breathing belly of a cat. Even those less inclined to bond with robots found appeal in such ideas, noting that a ``computer with a tail'' could make anthropomorphism tangible. These visions align with research on anthropomorphism and animism in object perception \cite{epley2007seeing}, as well as work on morphological interfaces that treat form and motion as expressive, material resources \cite{freyberg2023morphological}. Such distributed, embedded forms of emotional interaction also aligns with HCI research on enlivening everyday artifacts to support affective expression and communication \cite{ackema2004beyond}. By distributing partial morphologies (e.g., ears, tails, rhythmic breathing) across everyday objects, participants framed emotional interaction as ambient and situated within daily life.


\textbf{\textit{Design implication. Attach distributed animal-like features to everyday environments and objects to extend companionship beyond a single robot.}} This entails (1) embedding expressive morphologies into familiar objects, (2) tailoring movements to contextual cues (e.g., reminders, comfort, health), and (3) using distributed embodiments to create lightweight, ambient companionship. Such strategies expand the design space from isolated robots to ecologies of morphological expression embedded in daily life.

\subsection{Ethical and Relational Implications of Designing Robotic Pets}


Zoomorphic robotic pets reshape interaction dynamics at three levels: human–robot relationships, human–animal relationships, and multispecies cohabitation. Reflections from the workshop highlight ethical considerations across all three, outlining the broader moral ecology in which such companions operate.

\subsubsection{Human–Robot Relationships: Attachment, Agency, and Responsibility}
Participants imagined three broad styles of relating to zoomorphic companion robots (Section \ref{Relationship Styles}): companionship without obligation (\emph{responsibility-exempt}), reciprocal bonds that involved ``giving back'' (\emph{reciprocal-caring}), and flexible arrangements where responsibility could be turned on or off (\emph{controlled-boundary}). \Revision{These patterns suggest that robotic pets are granted a degree of \textit{moral standing} \cite{friedman2003hardware,kahn2012technological}, and that this can generate a potential \textit{moral burden} that some participants experienced as uncomfortable \cite{lowenthal2010social,darling2016extending}. Together, relationship patterns modeled on real animals can increase closeness but also evoke caregiving obligations, leading some users to withdraw to avoid this burden \cite{arnd2015sherry,coeckelbergh2011humans}. Features that simulate illness, aging, or death likewise introduce expectations of care and should be treated as intentional design decisions rather than neutral additions \cite{friedman2003hardware}.}

\subsubsection{Impacts on Human–Animal Relationships: Comparison, Substitution, and Ethical Reflection}

The workshop revealed several ways in which robotic pets may influence users’ relationships with real animals. Some participants realized that projecting ``perfect pet'' or highly compliant behaviors onto the robot made them reconsider expectations they had placed on their pets, especially when those expectations exceeded animals’ natural capacities (see Section~\ref{Robot Identity}). \Revision{Others viewed robotic pets as situational supplements, such as for workplaces where animals are not permitted, or as preparation for potential pet ownership rather than as replacements \cite{lazar2016why, krueger2021human, kahn2004robotic}. Despite this, prior work in HRI and ACI warns that the growing presence of robotic companions may shift how people evaluate or engage with living animals. Increased familiarity with predictable, always-responsive robots could alter people's expectations of emotional labor, behavioral consistency, or convenience in real animals \cite{sparrow2002march, sharkey2020crying}. These considerations suggest that the introduction of robotic pets may have downstream effects on human–animal relationships, requiring continued ethical attention as such systems become more common.}

\Revision{
\subsubsection{Robot–Animal Interactions: Cohabitation, Safety, and Animals as Stakeholders}

As robotic pets enter everyday settings, they will coexist with living animals in shared multispecies environments, making real animals stakeholders whose safety and comfort must be considered \cite{westerlaken2016becoming}. Prior ACI research shows that robot motion features, such as speed, trajectory, and proximity, can elicit unintended behavioral responses in animals \cite{schneiders2024designing,mancini2011animal}. Although MojiKit is designed for human use, future work should ensure that behaviors created with the toolkit do not cause distress to real animals in cohabited environments \cite{cucumak2025designing}. This requires grounding motion patterns in basic ethological principles and considering multispecies cohabitation in behavior design.
}

\subsection{Limitations \& Future Work}

While MojiKit revealed how animal-inspired robots can be designed for emotional interaction, several limitations remain. First, we did not conduct follow-up evaluations to test the effectiveness of participants’ designs, so findings reflect short-term exploration rather than sustained use. Second, the system has not been deployed in real-world settings, leaving the proposed interaction patterns untested in everyday contexts. Third, the prototype was technically constrained, lacking perceptual sensing and supporting only a limited set of animal-like features, which restricted realism and diversity of interactions. 
\Revision{
Fourth, our evaluation remained largely anthropocentric: we focused on human–robot interaction and did not consider how robot behaviors may affect, or be perceived by, living animals in shared environments. In addition, our workshop primarily recruited participants with design or development backgrounds, rather than the broader range of potential users of robotic pets.
}

Future work should address these gaps by (1) developing MojiKit into a long-term design toolkit with richer perception, morphology, and digitalized resources, (2) exploring perspectives across multi-user and cross-cultural groups, and (3) deploying outcomes in domains such as eldercare, education, and therapy to assess practical value and ethical implications. 
\Revision{
(4) Responding to calls for More-than-Human design, we plan to investigate interspecies interactions to understand how real animals respond to robot-generated behaviors and to ensure safe, ethical cohabitation. (5) Future studies will also recruit more diverse user populations, including older adults, children, and long-term pet owners, and conduct longitudinal evaluations to examine the durability and ecological validity of the designed behaviors.
}

\section{Conclusion}
This paper presented MojiKit, a design probe combining structured behavioral resources with a tangible prototype to support the design of emotionally expressive behaviors for animal-inspired companion robots. Nine co-creation workshops with 18 users showed how the tool supported structured pathways, lowered creative barriers, and sparked ideas beyond its current capabilities. We also observed how users’ orientations, shaped by pet experience and personal preferences, produced different strategies and trajectories in exploring affective behavior. Our study highlights both the opportunities and challenges of embedding structured resources into creative HRI, and suggests future directions including extending perceptual capabilities, exploring distributed animal-inspired features, and supporting long-term engagement.

\section{Disclosure about Use of LLM}
Large language models (LLMs) were used as supportive tools. GPT-4o via Cursor IDE assisted in drafting parts of the Behavior Control Studio’s web interface and Python scripts for quantitative analysis, with all code subsequently reviewed and adapted by the authors. GPT-5 was used to help translate and organize the workshop outcomes table and to draft teaser illustrations, which were then edited and finalized by the authors. All core research activities, including study design, data collection, analysis, and interpretation, were conducted independently by the authors.


\begin{acks}
This work was supported by the Guangdong Provincial Key Laboratory of Integrated Communication, Sensing and Computation for Ubiquitous Internet of Things (No. 2023B1212010007), the 111 Center (No. D25008), the Guangzhou City–University Joint Research Grant (GP2025M024), and the Key Project of the Institute of Software in Chinese Academy of Sciences (ISCAS-ZD-202401). 
We would also like to thank Shuyan Ren and Lveyang Zhang for their valuable assistance and support in the project.
\end{acks}

\bibliographystyle{ACM-Reference-Format}
\bibliography{0-reference}

\appendix

\section{Interview Protocol for Card Evaluation}\label{sec:cards design interview}

\subsection*{Warm-up \& Background}
\begin{itemize}
    \item Could you briefly introduce your relationship with your pet (e.g., how long you have kept a cat or dog, how frequently and in what contexts you usually interact)?
    \item How would you describe your level of observation and understanding of pet behaviors?
\end{itemize}

\subsection*{Overall Impression}
\begin{itemize}
    \item After browsing through this set of cards, what is your first impression?
    \item Do you find the cards easy to understand and use? Why or why not?
\end{itemize}

\subsection*{Organization \& Clarity}
\begin{itemize}
    \item Do you think the division of sections in the cards is clear and consistent with the logical flow of human–pet interaction?
    \item In practice, can you easily locate information related to a specific interaction scenario?
    \item Are there parts where the classification feels too complex or too simplistic?
\end{itemize}

\subsection*{Content Coverage}
\begin{itemize}
    \item Does the information in the cards cover the main aspects of your daily interactions with pets?
    \item Are there important situations or behaviors you feel are missing (e.g., specific environments or animal reactions)?
    \item Regarding the “emotion–behavior” mapping of animals, are there aspects you think need to be supplemented or revised?
\end{itemize}

\subsection*{Usability \& Readability}
\begin{itemize}
    \item If used as a design reference tool, do you think the cards can help designers better understand human–pet interactions?
    \item When consulting the cards, do you find the information easy to retrieve, or are there expressions that might cause confusion?
    \item In terms of wording, are there places that feel too academic or insufficiently intuitive?
\end{itemize}

\subsection*{Suggestions for Improvement}
\begin{itemize}
    \item Which parts should be further detailed or illustrated with examples?
    \item Which parts could be merged, simplified, or rephrased?
    \item Do you have ideas for additional categories or interaction scenarios that should be included?
\end{itemize}

\subsection*{Wrap-up}
\begin{itemize}
    \item What do you find most valuable about this set of cards?
    \item If you were to use it in your own design process, how would you apply it?
    \item Do you have any other suggestions or thoughts?
\end{itemize}

\section{Evaluation Scales and Interview Protocol} \label{Evaluation Scales and Interview Protocol in workshop}

This appendix provides the full materials used for data collection in our study. 
Section~\ref{sec:scales} lists the targeted evaluation scales administered after the workshop, including items for the Design Space and MojiKit usability. 
Section~\ref{sec:interview} presents the semi-structured interview guide used to capture participants’ design rationale, use of the Design Space, experiences with MojiKit, and overall reflections.

\subsection{Scale Items} \label{sec:scales}

\subsubsection{Targeted Evaluation of the Design Reference Cards}
\textbf{A. Information Retrieval Efficiency}
\begin{enumerate}
    \item I can quickly find the emotional or behavioral references I need.  
    \item The card structure makes it easy for me to locate information.  
\end{enumerate}

\textbf{B. Inspiration}
\begin{enumerate}
    \setcounter{enumi}{2}
    \item The cards provided me with new ideas I had not considered before.  
    \item The cards helped me expand the range of possible design directions.  
\end{enumerate}

\textbf{C. Comprehensibility}
\begin{enumerate}
    \setcounter{enumi}{4}
    \item The terminology and categories in the cards are easy to understand.  
    \item I can easily understand the relationships between different types of cards.  
\end{enumerate}

\subsubsection{MomoBot’s Support in the Creative Process} 
\textbf{A. Parameter Adjustment and Feedback Comprehension}
\begin{enumerate}
    \item MomoBot’s real-time feedback makes it easier for me to understand how parameters influence behaviors.  
    \item After adjusting parameters, I can clearly observe the changes in the robot’s actions.  
\end{enumerate}

\textbf{B. Idea Realization and Expression}
\begin{enumerate}
    \setcounter{enumi}{2}
    \item Even with limited actions, I can combine parameters to express the emotions I intend.  
    \item MomoBot’s functions are sufficient to support me in conceiving and realizing a complete interaction pattern.  
\end{enumerate}

\textbf{C. Creative Process Support}
\begin{enumerate}
    \setcounter{enumi}{4}
    \item MomoBot encourages me to try different combinations of actions and expressions.  
    \item Through MomoBot’s real-time feedback, I can form clearer perceptions and judgments of my design.  
\end{enumerate}

\textbf{D. Design Iteration and Adaptation}
\begin{enumerate}
    \setcounter{enumi}{6}
    \item Sometimes MomoBot cannot fully realize my ideas, but I am willing to adapt them by adjusting or simplifying actions.  
    \item Over time, I have gradually learned how to use MomoBot to design interaction patterns.  
\end{enumerate}

\subsection{Interview Guide}\label{sec:interview}

\textbf{Section 1: Design Thinking and Decision-Making Process}
\begin{enumerate}
    \item When designing an interaction pattern, which part do you usually begin with?  
      -- Defining the emotional goal / Setting the trigger / Imagining the robot’s action / Other  
    \item How did you decide on the emotional communication elements in your design? What references did you use?  
    \item When designing triggers (human actions or environmental factors), where did your inspiration come from?  
    \item What factors did you consider when designing the robot’s action combinations?  
      -- Similarity to real animal behavior  
      -- Degree of anthropomorphism  
      -- Feasibility within MojiKit  
    \item Did you revise your initial ideas during the process? Why?  
\end{enumerate}

\textbf{Section 2: Use of the Design Space and Sources of Ideas}
\begin{enumerate}
    \setcounter{enumi}{5}
    \item How did you use the Design Space cards in your creation?  
      -- For knowledge / For inspiration / To verify ideas / To refine action details  
    \item In what sequence did you usually consult the cards (e.g., human needs → triggers → robot actions), and which part did you use most often?  
    \item What were the main sources of your design ideas today?  
      -- Based on prior experience and impressions of real animal interactions  
      -- Generated after consulting the Design Space  
      -- Please provide examples for each case.  
    \item Did any content in the Design Space influence your interaction concepts? Please give examples.  
    \item Was there any missing information in the Design Space that made design difficult?  
    \item How could the cards be improved to be more useful?  
\end{enumerate}

\textbf{Section 3: Role of MojiKit in Design Realization}
\begin{enumerate}
    \setcounter{enumi}{12}
    \item Which parameters did you focus on when adjusting and combining actions in MojiKit?  
    \item Did MojiKit’s action performance affect your design of interaction patterns? Please give examples.  
    \item Did you encounter situations where your idea was good but MojiKit could not realize it? How did you handle this?  
    \item What additional functions would you like MojiKit to have?  
\end{enumerate}

\textbf{Section 4: Overall Creative Experience}
\begin{enumerate}
    \setcounter{enumi}{16}
    \item Which step in the process was most challenging for you? Why?  
    \item Which step gave you the greatest sense of achievement or enjoyment?  
    \item If you had more time, what improvements would you make to your design?  
\end{enumerate}

\textbf{Section 5: Open Questions}
\begin{enumerate}
    \setcounter{enumi}{19}
    \item Beyond today’s tools, what other methods could help you design emotional interaction patterns for animal-like robots?  
    \item What other potential applications do you see for this toolkit?  
\end{enumerate}

\section{Complete Card Set}\label{sec:8 cards}

\begin{figure*}
    \centering
    \includegraphics[width=0.75\linewidth]{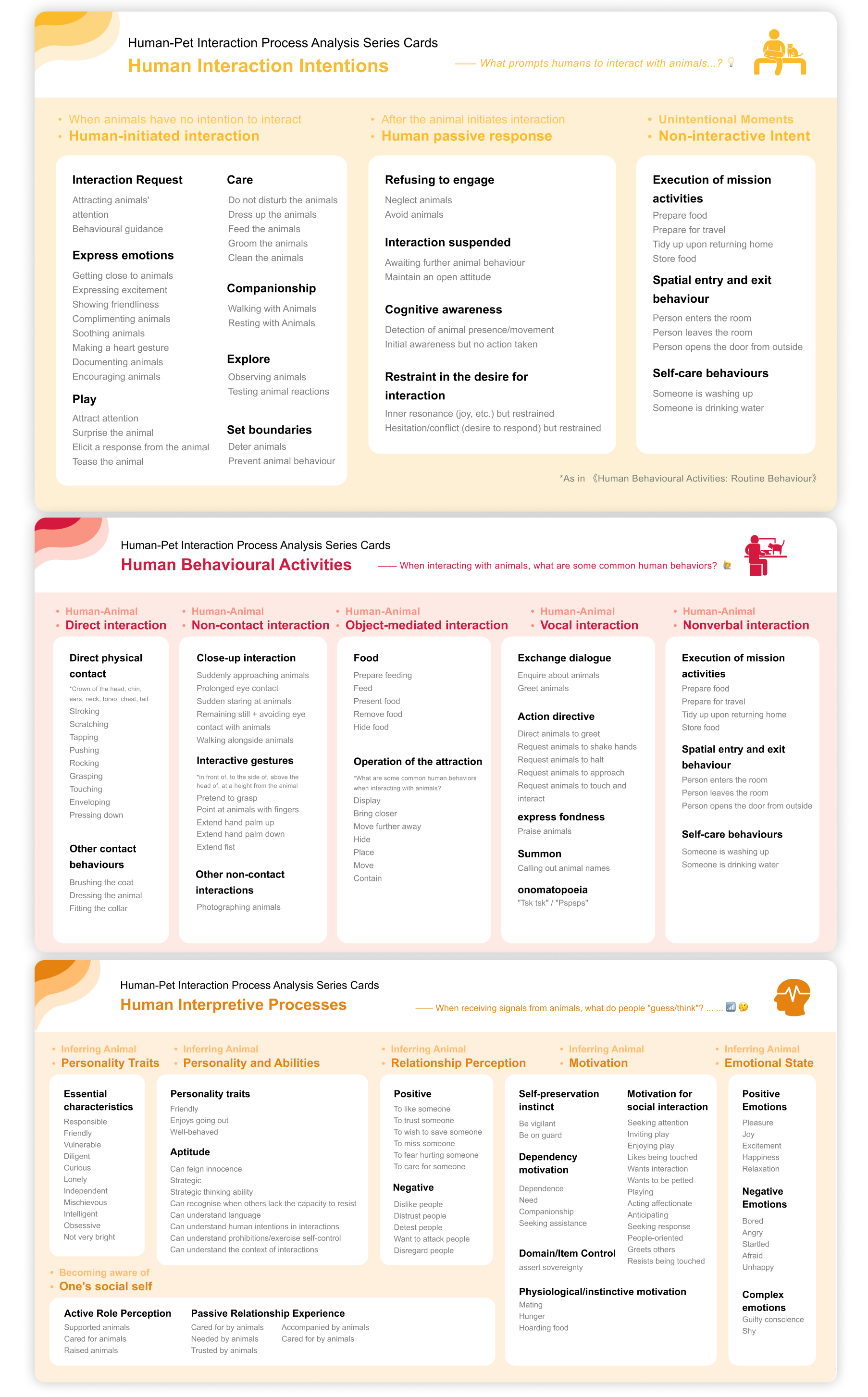}
    \caption{Human-Pet Interaction Process Analysis Series Cards: Human Interaction Intentions, Human Behavioural Activities, Human Interpretive Processes.}
    \Description{Figure showing three ``Human-centric'' cards from the Human-Pet Interaction Process Analysis Series Cards. The cards are stacked vertically. The top card is titled Human Interaction Intentions and organizes human-side intentions around interacting with animals into multiple labeled sections with example bullet points. The middle card is titled Human Behavioural Activities and categorizes observable human behaviors during human–animal situations (e.g., direct contact, non-contact, object-mediated, vocal, and nonverbal behaviors) in several column-like panels. The bottom card is titled Human Interpretive Processes and summarizes how people interpret animals’ signals, grouping interpretations such as inferred traits/abilities, relationship perception, motivation, and emotional state. All three cards share a consistent visual layout with headings and boxed lists. }
    \label{fig:cards-human}
\end{figure*}

\begin{figure*}
    \centering
    \includegraphics[width=0.65\linewidth]{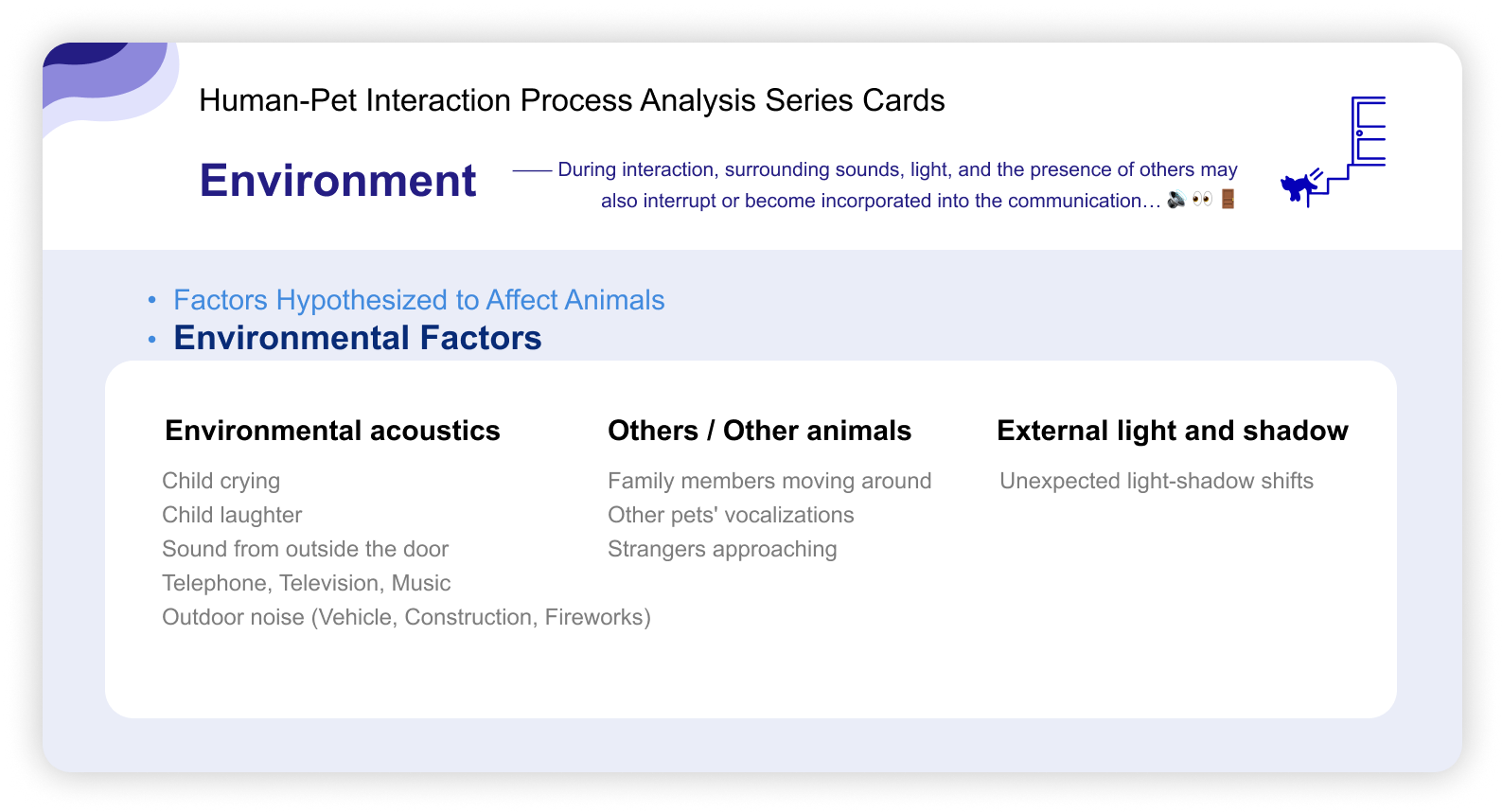}
    \caption{Human-Pet Interaction Process Analysis Series Cards: Environment.}
    \Description{The figure shows an ``Environment'' card from the Human–Pet Interaction Process Analysis Series Cards. The card focuses on environmental factors hypothesized to affect animals during interaction. It presents a structured list of environmental factors organized into three sections: environmental acoustics (such as human voices, household sounds, and outdoor noise), others or other animals (such as family members moving, vocalizations, or strangers approaching), and external light and shadow (such as unexpected changes in lighting). The information is arranged in labeled columns with example items listed under each category. }
    \label{fig:3-Environment}
\end{figure*}

\begin{figure*}
    \centering
    \includegraphics[width=0.65\linewidth]{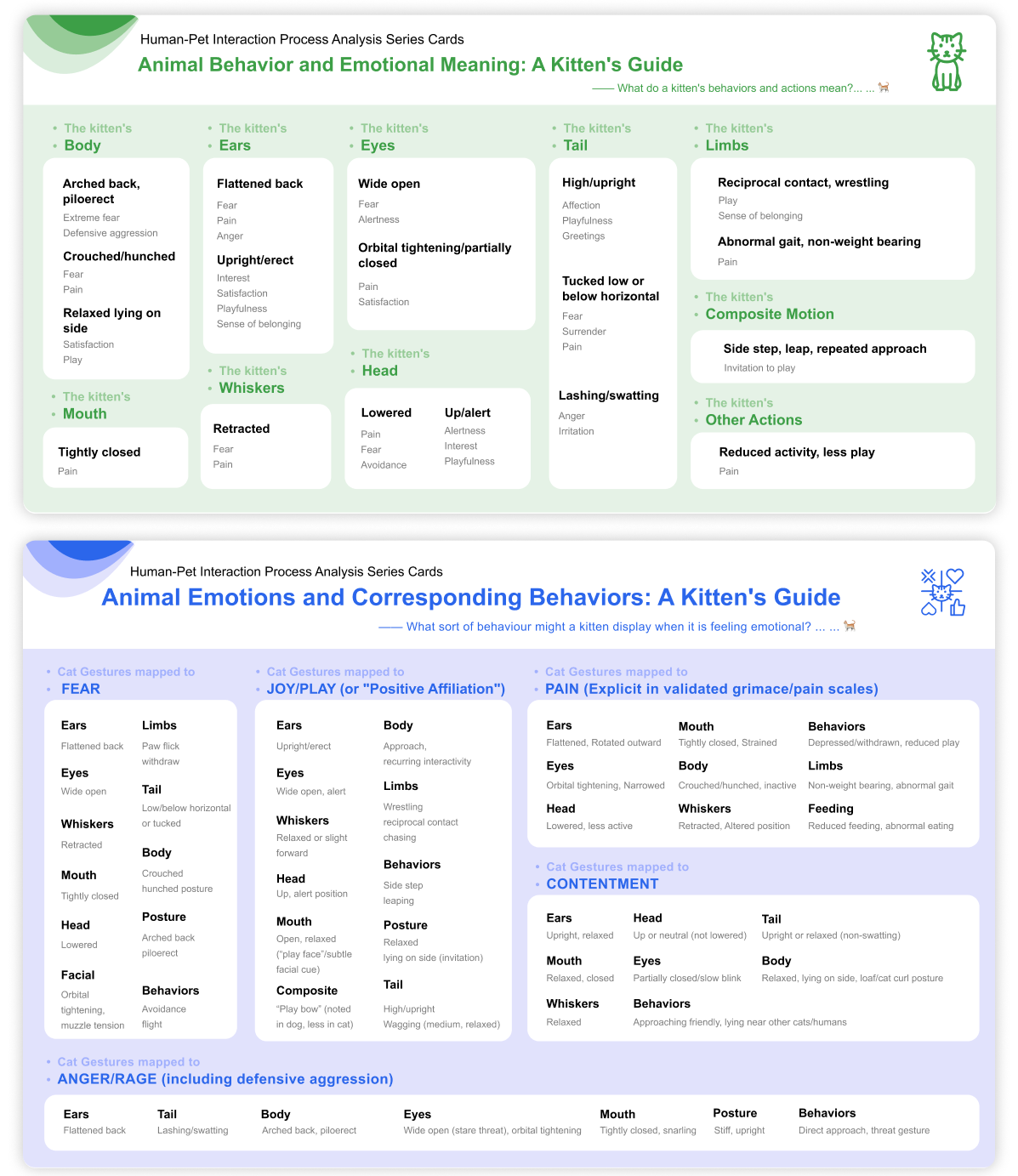}
    \caption{Human-Pet Interaction Process Analysis Series Cards: Animal Behavior and Emotional Meaning - A Kitten's Guide, Animal Emotions and Corresponding Behaviors.}
    \Description{The figure shows two ``Animal (Cat)'' cards from the Human–Pet Interaction Process Analysis Series Cards, displayed one above the other. The top card is titled Animal Behavior and Emotional Meaning: A Kitten’s Guide and organizes common kitten body cues and postures by body parts, including body posture, ears, eyes, tail, limbs, whiskers, mouth, head, and composite motion, with example behaviors and associated emotional meanings listed in labeled boxes. The bottom card is titled Animal Emotions and Corresponding Behaviors: A Kitten’s Guide and maps emotional states (such as fear, play/joy, pain, contentment, and anger) to observable kitten gestures across body parts and behaviors. Both cards use structured panels, headings, and bullet lists to present relationships between animal behaviors and emotional interpretations. }
    \label{fig:cards-animal-cat}
\end{figure*}

\begin{figure*}
    \centering
    \includegraphics[width=0.7\linewidth]{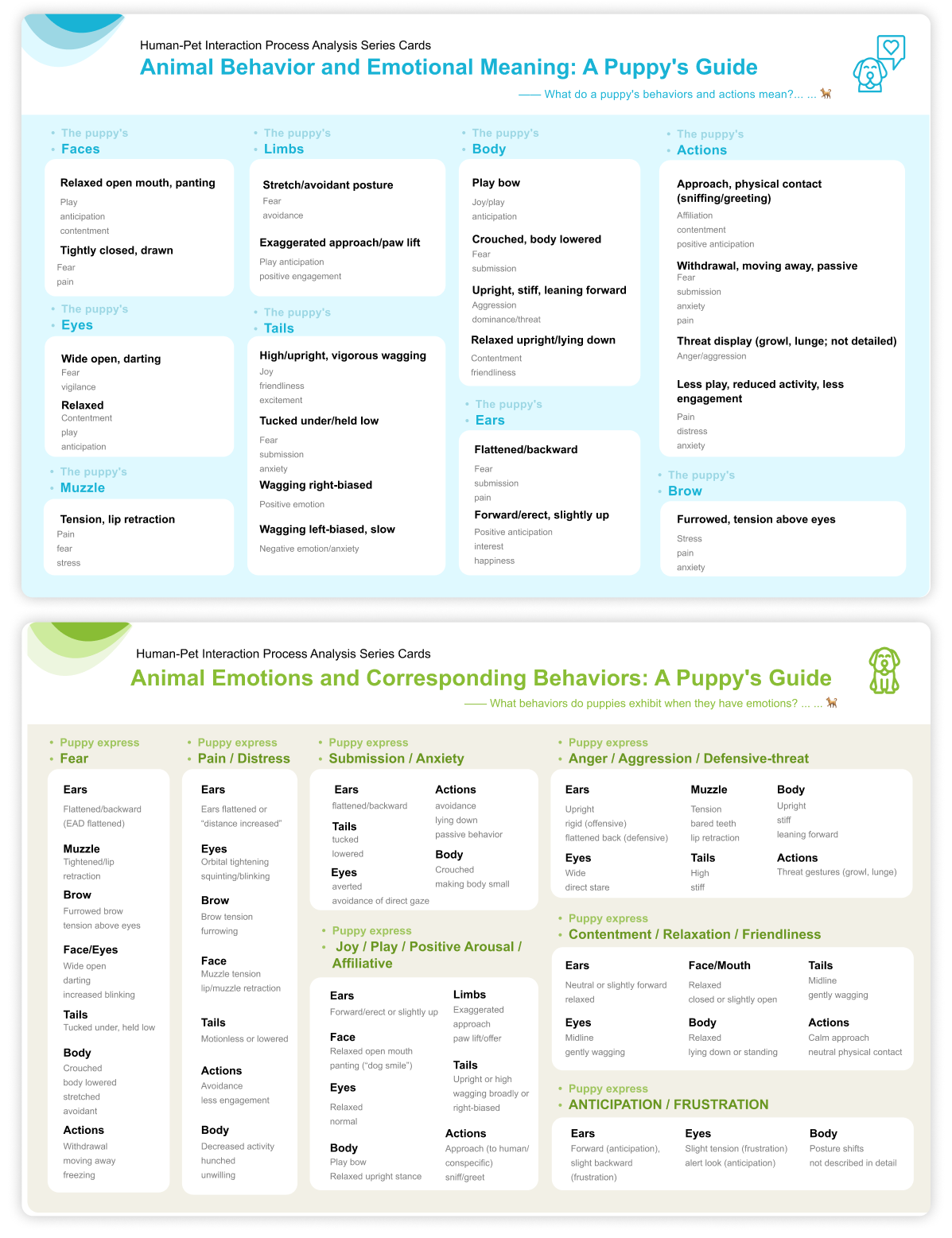}
    \caption{Human-Pet Interaction Process Analysis Series Cards: Animal Behavior and Emotional Meaning - A Puppy's Guide, Animal Emotions and Corresponding Behaviors.}
    \Description{The figure shows two ``Animal (Dog)'' cards related to puppies from the Human–Pet Interaction Process Analysis Series Cards, displayed one above the other. The top card is titled Animal Behavior and Emotional Meaning: A Puppy’s Guide and organizes common puppy body cues and actions by body parts, including face, eyes, muzzle, ears, tail, limbs, body posture, and actions, with example behaviors and associated emotional meanings listed in labeled sections. The bottom card is titled Animal Emotions and Corresponding Behaviors: A Puppy’s Guide and maps emotional states (such as fear, pain or distress, submission or anxiety, joy or play, anger or aggression, contentment, and anticipation) to observable puppy behaviors across body parts and actions. Both cards use structured panels, headings, and bullet lists to present relationships between animal behaviors and emotional interpretations. }
    \label{fig:cards-animal-dog}
\end{figure*}

This appendix provides the full set of eight design cards developed for the study.  
Each card includes structured observations of human–pet interactions, corresponding affective interpretations, and design implications for robot behavior.  
Figures \ref{fig:cards-human}, \ref{fig:3-Environment}, \ref{fig:cards-animal-cat} and \ref{fig:cards-animal-dog} present the cards in their original layout and visual design.  

\section{Interaction patterns collected from the workshop}\label{sec:design outcomes}
To complement the main text, Appendix Table \ref{tab:condensed_patterns} presents a condensed overview of the 35 validated interaction patterns generated in the workshop. Each entry summarizes the human intent, the corresponding robot behavior or feedback, and its affective meaning. 

\begin{table*}[]
\caption{Condensed summary of 35 validated interaction patterns, showing intent, robot behavior, and affective meaning.}
\Description{A comprehensive longtable summarizing 35 validated human-robot interaction patterns organized by participant group (G1-G9) and pattern ID. Three main columns show: (1) Human Intent categories (e.g., Attract attention, Desire for petting, Training, Companionship), (2) Corresponding Robot Behaviors (e.g., Head-turn toward sound, Rolling onto back, Tail wagging, Lying beside), and (3) Robot Affective Meanings (e.g., Responsive, Happy, Excited, Affectionate). Patterns demonstrate diverse interaction contexts, including greeting, play, discipline, and companionship. }
\label{tab:condensed_patterns}
\begin{tabular}{@{}llll@{}}
\toprule
\textbf{ID} & \textbf{Human Intent} & \textbf{Robot Behavior / Feedback} & \textbf{Robot Affective Meaning} \\ \midrule
G1-1 & Attract attention & Head-turn toward sound, return after confirmation & Responsive, attentive \\
G1-2 & Reject unwanted touch & Head shaking, trunk movement, leaving & Impatient, avoiding \\
G1-3 & Desire for petting & Roll onto back, chin raised, C-shape posture & Feeling needed, happy \\
G1-4 & Reward & Rush forward, cling with paws, licking & Affectionate gift, pampered \\
G2-1 & Respond to call & Tail wagging, walking over & Happy, eager to play \\
G2-2 & Companionship & Approaching, ear and head movement & Relaxed, positive \\
G2-3 & Sleeping companion & Lying beside & Dependence, security \\
G2-4 & Express need (mealtime) & Pawing at human & Independent, caring \\
G2-6 & Desire to play & Playful movement, toy interaction & Excited, affectionate \\
G2-7 & Happiness & Small vocal feedback & Caring, responsive \\
G3-1 & Cooperation & Call human, coordinate bug catching & Cooperative intent \\
G3-2 & Playfulness & Rolling on human’s body & Mischievous, amusing \\
G4-1 & Discipline & Stiff posture, quick retreat & Guilty, avoidant \\
G4-2 & First encounter & Slow eye opening, vocalizing, paw raise, head tilt & Curious, friendly \\
G4-3 & Affection & Slow approach, gentle tail wag, head lowering & Attachment, trust \\
G5-1 & Attention-seeking & Approach, upward gaze & Lonely, affectionate \\
G5-1 & Greeting & Tail wagging, hopping, barking & Excited, happy \\
G5-2 & Request to play & Bringing toy, wagging, circling & Friendly, expectant \\
G5-2 & Request companionship & Same as above & Lonely, expectant \\
G6-1 & Training & Sitting after repeated instruction & Learning, compliance \\
G6-2 & Daily companionship & Rolling, curled posture & Excited, relaxed \\
G6-3 & Greeting & Tail wagging, “dog smile,” circling & Affection, longing \\
G6-4 & Companionship & Approaching, lying beside & Secure, calm \\
G7-1 & Proactive attention-seeking & Limb banging on floor & Dissatisfied, demanding \\
G7-2 & Touch interaction & Head movement, ear twitching & Happy, responsive \\
G8-1 & Relaxation/comfort & Lying in arms, nuzzling, purring & Satisfied, secure \\
G8-2 & Intense affection & Flailing limbs, tail whipping & Alarmed, confused \\
G8-3 & Seeking companionship & Climbing desk, pushing cup & Lonely, attention-seeking \\
G8-4 & Energy release & Running, tail upright & Excited, energetic \\
G9-1 & Longing/affection & Running, nuzzling, vocalizing & Happy, attached \\
G9-2 & Bed companion & Lying on bed, nuzzling, ear movement & Affectionate, sleepy \\
G9-3 & Caring reminder & Patting bed & Affectionate, reminding \\
G9-4 & Companionship & Walking over, looking up & Happy, attentive \\
G9-5 & Companionship & Waiting outside bathroom, nuzzling & Affectionate, companionable \\ \bottomrule
\end{tabular}
\end{table*}

\end{document}